\documentclass[twocolumn]{aastex62}
\usepackage{graphicx}
\usepackage{natbib}
\usepackage{placeins}
\usepackage{amsmath}
\usepackage{tgcursor}
\usepackage{booktabs}
\usepackage{enumitem}
\setcitestyle{notesep={ }}

\begin{document}

\title{An evolutionary study of volatile chemistry in protoplanetary disks}

\author{Jennifer B. Bergner}
\affiliation{University of Chicago Department of the Geophysical Sciences, Chicago, IL 60637, USA}
\affiliation{NASA Sagan Fellow}

\author{Karin I. \"Oberg}
\affiliation{Center for Astrophysics $|$ Harvard \& Smithsonian, Cambridge, MA 02138, USA}

\author{Edwin A. Bergin}
\affiliation{Department of Astronomy, University of Michigan, Ann Arbor, MI 48109, USA}

\author{Sean M. Andrews}
\affiliation{Center for Astrophysics $|$ Harvard \& Smithsonian, Cambridge, MA 02138, USA}

\author{Geoffrey A. Blake}
\affiliation{Division of Geological \& Planetary Sciences, California Institute of Technology, Pasadena, CA 91125, USA}

\author{John M. Carpenter}
\affiliation{Joint ALMA Observatory, Alonso de C{\'o}rdova 3107 Vitacura, Santiago, Chile}

\author{L. Ilsedore Cleeves}
\affiliation{Department of Astronomy, University of Virginia, Charlottesville, VA 22904, USA}

\author{Viviana V. Guzm{\'a}n}
\affiliation{Instituto de Astrof{\'i}sica, Pontf{\'i}ficia Universidad Cat{\'o}lica de Chile, Av. Vicu{\~n}a Mackenna 4860, 7820436 Macul, Santiago, Chile}

\author{Jane Huang}
\affiliation{Center for Astrophysics $|$ Harvard \& Smithsonian, Cambridge, MA 02138, USA}

\author{Jes K. J{\o}rgensen}
\affiliation{Niels Bohr Institute \& Centre for Star and Planet Formation, University of Copenhagen, {\O}ster Voldgade 5-7, DK-1350 Copenhagen K., Denmark}

\author{Chunhua Qi}
\affiliation{Center for Astrophysics $|$ Harvard \& Smithsonian, Cambridge, MA 02138, USA}

\author{Kamber R. Schwarz}
\affiliation{Lunar and Planetary Laboratory, University of Arizona, Tucson, AZ 85721, USA}
\affiliation{NASA Sagan Fellow}

\author{Jonathan P. Williams}
\affiliation{Institute for Astronomy, University of Hawai'i at M$\bar{a}$noa, Honolulu, HI 96822, USA}

\author{David J. Wilner}
\affiliation{Center for Astrophysics $|$ Harvard \& Smithsonian, Cambridge, MA 02138, USA}

\begin{abstract}
\noindent The volatile composition of a planet is determined by the inventory of gas and ice in the parent disk.  The volatile chemistry in the disk is expected to evolve over time, though this evolution is poorly constrained observationally.  We present ALMA observations of C$^{18}$O, C$_2$H, and the isotopologues H$^{13}$CN, HC$^{15}$N, and DCN towards five Class 0/I disk candidates.  Combined with a sample of fourteen Class II disks presented in \citet{Bergner2019b}, this data set offers a view of volatile chemical evolution over the disk lifetime.  Our estimates of C$^{18}$O abundances are consistent with a rapid depletion of CO in the first $\sim$0.5--1 Myr of the disk lifetime.  We do not see evidence that C$_2$H and HCN formation are enhanced by CO depletion, possibly because the gas is already quite under-abundant in CO.  Further CO depletion may actually hinder their production by limiting the gas-phase carbon supply.  The embedded sources show several chemical differences compared to the Class II stage, which seem to arise from shielding of radiation by the envelope (impacting C$_2$H formation and HC$^{15}$N fractionation) and sublimation of ices from infalling material (impacting HCN and C$^{18}$O abundances).  Such chemical differences between Class 0/I and Class II sources may affect the volatile composition of planet-forming material at different stages in the disk lifetime.
\end{abstract}


\section{Introduction}
\label{sec:intro}
Planet formation takes place within the gas- and dust-rich disks surrounding young stars.  The volatile composition of a planet is set by the reservoirs of gas and ice in the disk at the time of its assembly.  It is therefore important to understand how the gas and ice compositions evolve over the disk lifetime.  In particular, the relative abundances of volatile elements (C/N/O) in an exoplanet's atmosphere offer a promising way to infer its formation history \citep[e.g.][]{Oberg2011, Cridland2016}, provided that we understand the chemistry of these elements in the disk.  Additionally, C, N, and O are the main ingredients of organic chemistry, and the inventories of these elements incorporated into nascent planets may have important implications for the viability of prebiotic chemistry.

One key trend to emerge from observations of volatiles in disks is that CO and H$_2$O appear depleted in the molecular layer relative to ISM abundances \citep{Dutrey2003, Chapillon2008, Hogerheijde2011, Favre2013, Bergin2013, Williams2014, McClure2016, Cleeves2016, Schwarz2016, Du2017}.  This is thought to reflect the transport of ice-covered dust grains to the disk midplane, where chemical and/or physical processes prevent the volatiles from re-entering the gas \citep[e.g.][]{Meijerink2009, Bergin2010, Krijt2016, Schwarz2018}.  The different volatilities and chemical reactivities of major volatile carriers (H$_2$O, CO, and N$_2$) mean that each element will be depleted to a different extent, with oxygen being most depleted, followed by carbon and then nitrogen.  Modeling of the CO, C$_2$H, and HCN emission in the IM Lup disk supports this depletion pattern \citep{Cleeves2018}. 

Besides impacting the relative distribution of CO and H$_2$O between the disk midplane and atmosphere, this depletion may also alter the chemistry of other volatiles in the disk atmosphere.  In particular, models predict that an O-depleted gas with a strong UV radiation field should facilitate the production of cyanides and hydrocarbons \citep{Du2015}.  Bright cyanide (CN, HCN, CH$_3$CN, HC$_3$N) and hydrocarbon (C$_2$H, C$_3$H$_2$) emission towards a number of protoplanetary disks has been interpreted as an outcome of this chemistry \citep{Bergin2016, Guzman2017, Bergner2018}.  

While this framework can explain the emission patterns in a few individual sources, we are lacking a systematic view of how the volatile chemistry evolves with time.  Depletion factors measured across different disks vary dramatically \citep[generally factors of a few to 100 for CO and two to four orders of magnitude for O traced by H$_2$O, e.g.][]{McClure2016, Du2017}, and it is unclear whether this is an evolutionary effect or a reflection of different depletion efficiencies in different physical environments.  Similarly, it is not clear at what stage volatile depletion becomes important for regulating disk chemistry.  The low CO abundance inferred for the $\sim$1 Myr IM Lup disk \citep[20$\times$ lower than ISM levels; ][]{Cleeves2018} suggests that depletion must, in at least some cases, occur very early in the disk lifetime.  

The physical evolution of a protostar begins with an object deeply embedded in its envelope (termed Class 0).  The envelope undergoes infall and accretion onto a circumstellar disk (Class I) and is eventually dispersed, leaving only a protoplanetary disk (Class II).  Classes are generally assigned based on the infrared slope of the SED and/or the protostar's bolometric temperature \citep{Dunham2014}.  Recent evidence indicates that grain growth and possibly planet formation are well underway by the Class II stage \citep{Andrews2018, Zhang2018}, and perhaps already by the embedded stage \citep{ALMA2015, Harsono2018}.  This implies that much of the physics and chemistry important for planet formation may be set in the Class 0/I (protostellar) disk stage.  

It has only recently become possible to disentangle Class 0/I disks from the envelope using high spatial resolution observations of line kinematics \citep[e.g.][]{Murillo2013,Ohashi2014}, and our understanding of how the protostellar disk chemistry relates to protoplanetary disk chemistry is poor.  Notably, the degree of volatile depletion in protostellar disks is unclear.  \citet{Anderl2016} see evidence for $\sim$1 order of magnitude CO depletion in a sample of Class 0 envelopes, and \citet{Harsono2020} find low water abundances in embedded Class I disks.  Meanwhile, \citet{vantHoff2018} and \citet{Zhang2020} find evidence for an ISM CO abundance in the embedded disks L1527, TMC 1A, HL Tau, and DG Tau.  Additional observations  of embedded sources, particularly those that resolve the protostellar core or disk from the envelope, are needed to provide insight into the chemistry and depletion of volatile molecules in protostellar disks.

Further constraint on the volatile chemistry in disks is offered by molecular isotopologues.  Different isotopic fractionation pathways are active under specific physical conditions, and so studying fractionation can provide insight into what types of chemistry are active at different evolutionary stages.  This is especially important for interpreting fractionation patterns seen in Solar system bodies, which are often used to infer where in the Solar nebula and from what material different objects formed \citep[e.g][]{Kerridge1987, Messenger1997, Ehrenfreund2000}.

The HCN isotopologues DCN and HC$^{15}$N are bright at mm wavelengths and relatively abundant throughout the star formation sequence, offering a valuable probe of deuterium and $^{15}$N fractionation.  Generally, D fractionation is expected to proceed by gas-phase ion exchange reactions that favor the incorporation of D into molecules in cold environments.  Fractionation pathways beginning with H$_2$D$^+$ are predicted to be efficient below 30 K \citep[e.g][]{Millar1989, Pagani1992}.  Pathways beginning with CH$_2$D$^+$ were originally expected to be efficient up to 80 K and were more recently predicted to be active up to 300 K \citep{Roberts2000, Roueff2013, Favre2015}.  CH$_2$D$^+$ initiation was expected to dominate DCN formation in early chemical models \citep{Millar1989}, though modeling by \citet{Willacy2007} introduced pathways to form DCN starting with H$_2$D$^+$ or D$_3^+$, both of which should favor colder environments.  Though smaller in magnitude than D fractionation, $^{15}$N fractionation also proceeds through gas-phase ion exchange reactions in cold environments \citep{Terzieva2000, Rodgers2008}.  Additionally, in strongly UV-irradiated environments such as disk atmospheres, selective dissociation of $^{15}$N$^{14}$N over N$_2$ due to different self-shielding efficiencies also enhances the incorporation of $^{15}$N into molecules besides N$_2$ \citep{Heays2014}.  

Recently, DCN/HCN and HC$^{15}$N/HCN fractionation patterns were surveyed in a sample of six Class II disks by \citet{Huang2017} and \citet{Guzman2017}.  \citet{Huang2017} find high DCN/HCN ratios in disks, consistent with measurements in pre- and proto-stellar sources, but with a wide morphological diversity suggestive of an in situ chemistry dependent on the unique physical conditions in each disk.  \citet{Guzman2017} find high HC$^{15}$N/HCN ratios consistent with values measured in cold gas and in comets, suggestive of inheritance.  Still, for the one disk with spatially resolved emission (V4046 Sgr), an increasing HCN/HC$^{15}$N ratio moving outwards in the disk suggests that in situ photodissociation-driven fractionation is important.  A similar result is seen in the TW Hya disk by \citet{Hily-Blant2019}.  Measurements of D and $^{15}$N fractionation in HCN towards Class 0/I disks are still needed to evaluate whether younger disks demonstrate similar fractionation patterns, or if there is a distinct Class II fractionation chemistry.

In this work, we present an evolutionary study of volatile chemistry in Class 0, I, and II sources.  We focus our analysis on two chemical regimes: (i) the relationship between hydrocarbon, cyanide, and CO chemical families as traced by C$_2$H, HCN, and C$^{18}$O, and (ii) $^{15}$N/$^{14}$N fractionation and D/H fractionation in HCN.  Our source targets consist of five protostellar (Class 0/I) disk candidates and fourteen Class II disks.  C$^{18}$O, C$_2$H, H$^{13}$CN, DCN, and HC$^{15}$N observations towards the Class 0/I sources are presented in this work, along with HCN isotopologue observations for a subset of the Class II sources.  C$_2$H, HCN, and C$^{18}$O observations towards the Class II disks were previously presented in \citet{Bergner2019b}.  HC$^{15}$N and DCN were also previously targeted towards six disks in our sample in \citet{Guzman2017} and \citet{Huang2017}.  Combining these data sets, we can test the paradigm of volatile depletion and its chemical side-effects, explore isotope fractionation chemistry, and distinguish the extent to which Class II disks represent a distinct chemical regime.  In Section \ref{sec:obs} we describe our source samples and observations.  Section \ref{sec:obs_results} presents the observational results, including line detections and emission morphologies.  In Section \ref{sec:cd_abunds} we derive column density and abundance estimates, and explore evolutionary trends in the C$_2$H, HCN, and C$^{18}$O abundances.  In Section \ref{sec:hcn} we derive HCN/HC$^{15}$N and HCN/DCN ratios across the source sample.  Lastly, in Section \ref{sec:disc} we discuss trends in the C$_2$H, HCN, and C$^{18}$O chemistries and in HCN/HC$^{15}$N and HCN/DCN ratios, and implications for the volatile chemical evolution over the disk lifetime.

\begin{deluxetable*}{lcl}[h]
	\tabletypesize{\footnotesize}
	\tablecaption{Summary of previously unpublished observations \label{tab:obs_summary}}
	\tablecolumns{3} 
	\tablewidth{\textwidth} 
	\tablehead{
		\colhead{Transition}                          &
		\colhead{ALMA Project Code}                               & 
		\colhead{Source targets}                     
}                           
\startdata
CO 2--1 & 2015.1.00964.S & Ser-emb 1, 7, 8, 15, 17 \\
$^{13}$CO 2--1 & 2015.1.00964.S & J1609, J1612, J1614 \\
C$^{18}$O 2--1 & 2015.1.00964.S & Ser-emb 1, 7, 8, 15, 17 \\
C$_2$H  N=3--2, J=$\frac{5}{2}$--$\frac{5}{2}$, F=3--3 & 2015.1.00964.S & Ser-emb 1, 7, 8, 15, 17 \\
HCN 3--2 &  2015.1.00964.S & HD 143006, J1604, J1609, J1612, J1614 \\
H$^{13}$CN 3--2 &  2015.1.00964.S & Ser-emb 1, 7, 8, 15, 17 \\ 
 & 2016.1.00627.S & CI Tau, DM Tau, DO Tau, LkCa 15, MWC 480 \\
 HC$^{15}$N 3--2 & 2015.1.00964.S & Ser-emb 1, 7, 8, 15, 17 \\ 
  & 2016.1.00627.S & CI Tau, DM Tau, DO Tau, LkCa 15, MWC 480 \\
 DCN 3--2 & 2015.1.00964.S & Ser-emb 1, 7, 8, 15, 17; HD 143006, J1604, J1609, J1612, J1614 \\ 
  & 2016.1.00627.S & CI Tau, DM Tau, DO Tau \\
\enddata
\tablenotetext{}{}
\end{deluxetable*}

\begin{deluxetable*}{llllllll} 
	\tabletypesize{\footnotesize}
	\tablecaption{Source summary \label{tab:sourcedat}}
	\tablecolumns{8} 
	\tablewidth{\textwidth}     
	\tablehead{\multicolumn{8}{c}{Class 0/I targets$^{a,b}$}}                
\startdata
Source & Region & R.A. (J2000) & Dec. (J2000) & Class & $T_{bol}$ (K) & $M_{env}$ ($M_\odot$) & $L_{bol}$ ($L_\odot$) \\
 \hline 
Ser-emb 1$^*$ & Serpens &18:29:09.1 & 0:31:30.9 & 0   & 39 [2]     & 3.1 [0.05]  & 4.1 [0.3]           \\
Ser-emb 7$^*$ & Serpens & 18:28:54.1 & 0:29:30.0 & 0   & 58 [13]   & 4.3 [0.4]    & 7.9 [0.3]          \\
Ser-emb 8$^*$ & Serpens & 18:29:48.1 & 1:16:43.7 & 0   & 58 [16]   & 9.4 [0.3]    & 5.4$^c$ [6.2]   \\
Ser-emb 15$^*$ & Serpens & 18:29:54.3 & 0:36:00.8 & I & 101 [43] &1.3 [0.1]    &  0.4 [0.6]         \\
Ser-emb 17$^*$ & Serpens & 18:29:06.2 & 0:30:43.1 & I & 117 [21]  & 3.6 [0.4]   & 3.8 [3.3]          \\
\hline 
\multicolumn{8}{c}{Class II targets} \\
\hline
Source & Region & R.A. (J2000) & Dec. (J2000) & Dist.$^d$ (pc) & Age$^e$ (Myr) & $M_\star$ $^e$ ($M_\odot$) & $L_\star$ $^e$ ($L_\odot$)\\
\hline 
CI Tau$^*$         & Taurus & 04:33:52.0 & 22:50:29.8 & 158.7 &  0.7 (0.4--1.8) & 0.66 & 1.20 \\
DM Tau$^*$        &Taurus & 04:33:48.7 & 18:10:9.7 & 145.1 & 4.0 (2.5--7.1) & 0.53 & 0.24 \\
DO Tau$^*$        &Taurus & 04:38:28.6 & 26:10:49.1 & 139.4 & 0.4 (0.1--0.9) & 0.45 & 1.40 \\
HD 143006$^*$  &Upper Sco & 15:58:36.9 & -22:57:15.5 & 166.1 & 4.0 & 1.78 & 3.80 \\
J1604-2130$^*$ &Upper Sco & 16:04:21.6 & -21:30:28.9 & 150.1 & 13.8 (7.4-32) & 1.11 & 0.62 \\
J1609-1908$^*$ &Upper Sco & 16:09:00.7 & -19:08:53.1 & 137.6 & 4.8 (2.5--9.1) & 0.68 & 0.32 \\
J1612-1859$^*$ &Upper Sco & 16:12:39.2 & -18:59:28.9 & 139.1 & 4.1 (1.8--8.3) & 0.56 & 0.29 \\
J1614-1906$^*$ &Upper Sco & 16:14:20.3 & -19:06:48.5 & 143.9 & 2.1 (1.0--5.4) & 0.60 & 0.46 \\
LkCa 15$^*$       &Taurus & 04:39:17.8 & 22:21:03.1 & 158.9 & 2.0 (0.9--4.3) & 1.03 & 1.04 \\
MWC 480$^*$    &Taurus & 04:58:46.3 & 29:50:36.6 & 161.8 & 6.5 & 1.84 & 25 \\
AS 209               &Ophiuchus & 16:49:15.3 & -14:22:09.0 &121.0 & 1.0 & 0.83 & 1.41 \\
HD 163296        &Isolated? & 17:56:21.3 & -21:57:22.5 & 101.5 & 13 & 2.04 & 17 \\
IM Lup               & Lupus& 15:56:09.2 & -37:56:06.5 & 158.4 & 0.5 & 0.89 & 2.57 \\
V4046 Sgr         & $\beta$ Pic & 18:14:10.5 & -32:47:35.3 & 72.4 & 13 & 1.75 & 0.86 \\ 
\enddata
\tablenotetext{}{$^a$ Class 0/I source properties are taken from \citet{Enoch2009} and \citet{Enoch2011}. \\
$^b$ A distance of 436 pc is assumed for all Serpens sources \citep{Ortiz2018}. \\
$^c$ Likely an underestimate due to saturation in the 70$\mu$m flux measurement. \\
$^d$ From \textit{Gaia} Data Release 2 \citep{Gaia2018} \\
$^e$ Stellar properties are taken from \citet{Pegues2020} (calculated using the methodology outlined in \citet{Andrews2018}, except for V4046 Sgr which is taken from \citet{Rosenfeld2012}).  When available, age ranges reflecting the uncertainty are listed in parentheses. \\
$^*$ Indicates a source with new observations presented in this work.}
\end{deluxetable*}

\section{Observations}
\label{sec:obs}

The observations used in this project were taken as part of ALMA projects 2015.1.00964.S and 2016.1.00627.S (PI: K. \"Oberg).  The targets include protostellar (Class 0/I) disk candidates as well as protoplanetary (Class II) disks.  We aim to characterize evolutionary trends in (i) the CO, hydrocarbon, and cyanide chemistry, and (ii) deuterium and $^{15}$N fractionation in HCN.  We therefore focus on transitions of the CO isotopologues CO, $^{13}$CO, and C$^{18}$O; the HCN isotopologues HCN, H$^{13}$CN, HC$^{15}$N, and DCN; and C$_2$H.  Table \ref{tab:obs_summary} summarizes the previously unpublished molecular line observations within each ALMA program.  We note that this data set is complementary to that presented in \citet{Bergner2019b}, which included observations of CO or C$^{18}$O, HCN or H$^{13}$CN, and C$_2$H towards each Class II source.  In the following sections, we describe in more detail the target samples and ALMA observations for the Class 0/I and Class II groups.

\subsection{Protostellar (Class 0/I) disk candidates}
\label{sec:0I_sample}
The sample of protostellar disk candidates consists of five low-mass protostars in the Serpens cluster, described in Table \ref{tab:sourcedat}.  Each was identified as hosting a candidate disk based on non-zero flux at $>$50 k$\lambda$ $uv$ distances, corresponding to physical distances $<$1800 AU, in 230 GHz continuum observations \citep{Enoch2011}.  Ser-emb 1, 7, and 8 are classified as Class 0 sources, and Ser-emb 15 and 17 as Class I \citep{Enoch2009}.  \citet{Enoch2009} estimate that the Class 0 lifetime in Serpens is about 0.2 Myr, assuming a Class I lifetime of 0.54 Myr as found in \citet{Evans2009}.  A study of the organic molecule emission towards this source sample revealed that Ser-emb 1, Ser-emb 8, and Ser-emb 17 host hot corinos based on the detection of a warm, rich organic chemistry in the protostellar core \citep{Bergner2019c}.  

Full observational details can be found in \citet{Bergner2019c}.  Briefly, targets were observed by ALMA as part of project 2015.1.00964.S from May to June of 2016.  Two Band 6 spectral setups (217--233 GHz and 243--262 GHz) were covered.  This project makes use of spectral windows covering transitions of CO, C$^{18}$O, C$_2$H, H$^{13}$CN, HC$^{15}$N, and DCN.  All spectral windows have a channel width of 122 kHz, corresponding to $\sim$0.14--0.17 km/s across the range of frequencies.  

\subsection{Class II disks}
The full Class II disk sample is described in \citet{Bergner2019b}, and source properties are shown in Table \ref{tab:sourcedat}.  The sample spans a range of physical properties including age (0.4--14 Myr), stellar mass (0.5--2 M$_\odot$), and stellar luminosity (0.2--25 L$_\odot$).   In addition to the C$_2$H, HCN, and C$^{18}$O observations presented in \citet{Bergner2019b}, here we include previously unpublished observations of DCN, HC$^{15}$N, and $^{13}$CO lines towards a subset of the full disk sample.

The DCN 3--2 transition was observed towards HD 143006, J1604, J1609, J1612, and J1614 as part of ALMA project 2015.1.00964S; and towards CI Tau, DM Tau, and DO Tau as part of project 2016.1.00627.S.  The HC$^{15}$N 3--2 transition was observed towards CI Tau, DM Tau, DO Tau, LkCa 15, and MWC 480 as part of project 2016.1.00627.S.  For sources with DCN or HC$^{15}$N observations, we also re-analyze the HCN and H$^{13}$CN 3--2 observations originally presented in \citet{Bergner2019b} to facilitate a consistent analysis of isotopic ratios.  Lastly, we include previously unpublished observations of $^{13}$CO 2--1 towards J1609, J1612, and J1614, for which C$^{18}$O emission was not detected.  Observations took place from May--June of 2016 for project 2015.1.00964S and in December 2016 for project 2016.1.00627.S.  In both cases, two Band 6 spectral setups covered similar frequency ranges and spectral resolutions as described for the Serpens observations.

\subsection{Data reduction}
\label{sec:datared}

Following pipeline calibration of the ALMA data, we performed self-calibration using the combined continuum emission from spectral windows within a given sideband.  Depending on the continuum brightness, up to two rounds of phase self-calibration were performed.  Solution intervals ranged from 12.1--42.35 sec for the Class 0 sources, 6.05--12.1 sec for the Class I sources, 60.5--84.7 sec for the Class II Upper Sco sources, and 6.05--48.4 sec for the remaining Class II sources.  Continuum imaging was performed using Briggs weighting with a robust factor of 0.5.  The self-calibration solutions were applied to the spectral line data following continuum subtraction in the uv plane.  Calibrated and continuum-subtracted data were CLEANed to a 4$\sigma$ noise threshold in CASA 5.5.0 using Briggs weighting and the automasking task \texttt{auto-multithresh} in \texttt{tclean}.  We applied the standard automasking parameters for short baseline 12m line data (sidelobethreshold = 2.0, noisethreshold = 4.5, minbeamfrac = 0.3, negativethreshold = 15.0, lownoisethreshold = 1.5).  Automasking results were verified by comparing to hand-masked images for select lines including extended and compact emission morphologies, and bright and weak emission strengths.  For the Class 0/I sources, images were generated with a channel width of 0.25 km s$^{-1}$.  We used a robust factor of 0.5 for all molecules except C$_2$H, for which we used a robust factor of 2.0 due to its weak and diffuse emission.  For the Class II sources, imaging was performed with a channel width of 0.5 km s$^{-1}$.  A robust factor of 2.0 was used for all disks except for HD 143006 and J1604, whose bright emission permitted robust factors of 1.0 and 0.0 respectively. 

\section{Observational results}
\label{sec:obs_results}
\subsection{Class 0/I sources}
\subsubsection{Line detections and velocity structures}
\label{sec:class0_structures}
\begin{figure*}
	\includegraphics[width=\linewidth]{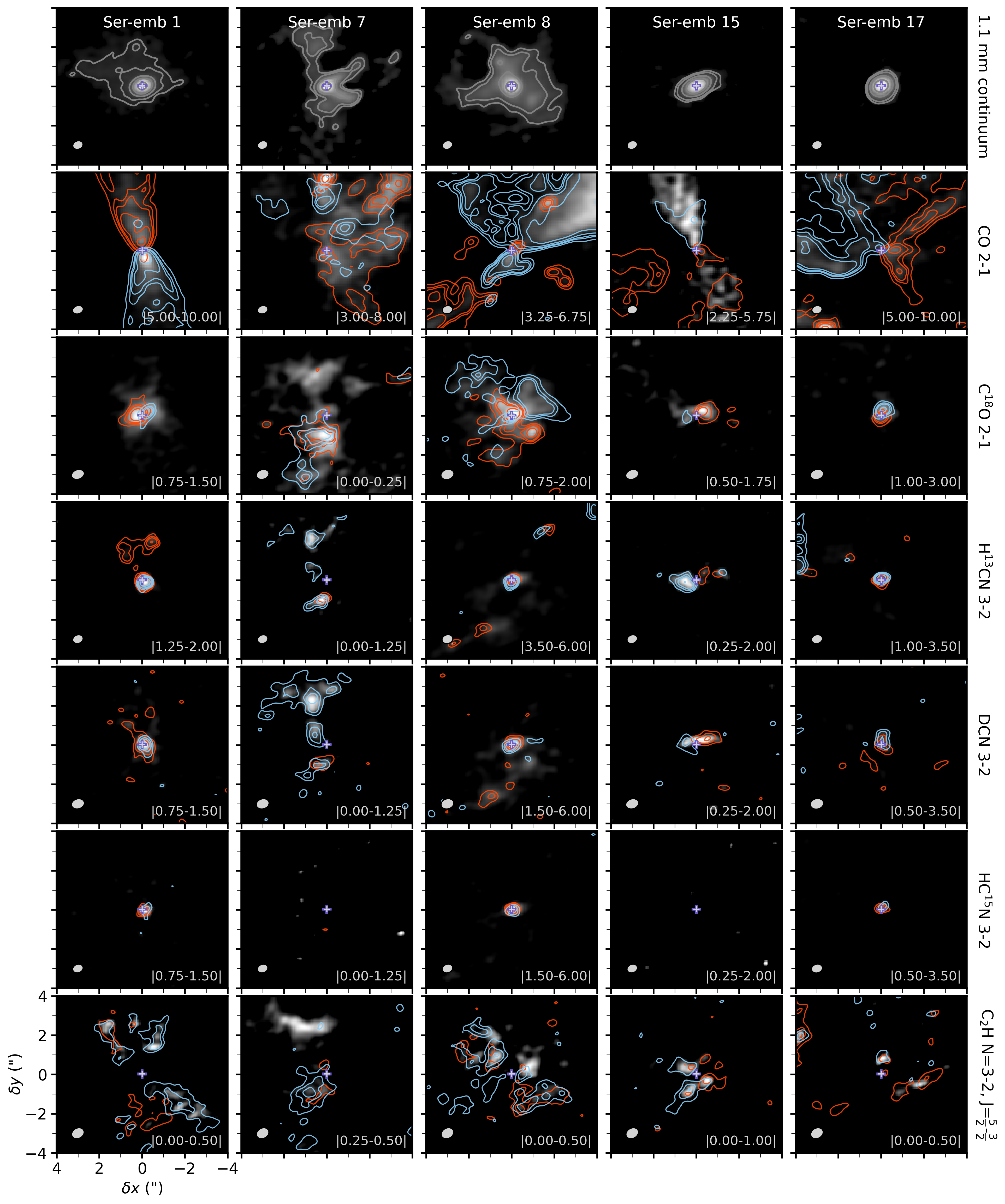}
	\caption{Protostellar disk candidate source structures (arranged with increasing $T_{bol}$ left to right).  The first row shows the 260 GHz continuum maps.  To show extended structure a power-law color scale is used with $\gamma$=0.3.  Contour levels correspond to [6,10,30,100,400]$\sigma$.  All other rows show integrated intensity maps for each molecular line in greyscale.  Colored contours correspond to the blue- and red-shifted emission within the velocity ranges indicated in the lower right of each panel (in km/s).  Contours correspond to [7,15,20,30,50]$\sigma$ for CO, [5,8,10,15]$\sigma$ for C$^{18}$O, [4,6,8,12]$\sigma$ for H$^{13}$CN and HC$^{15}$N, and [3,5,7,10]$\sigma$ for DCN and C$_2$H.  Moment 0 rms values for each velocity range can be found in Table \ref{tab:class0i_fluxes}.  Restoring beams are shown in the bottom left of each panel.  Color scales are normalized to each individual image, and emission below 3$\sigma$ is not shown.}
\label{fig:serp_vel_summary}
\end{figure*}

Figure \ref{fig:serp_vel_summary} shows an overview of the continuum and line emission observed towards the five Serpens protostellar disk candidates.  As discussed in \citet{Bergner2019c}, the continuum emission is more extended in the Class 0 sources (Ser-emb 1, 7, and 8) than in the Class I sources (Ser-emb 15 and 17), reflecting that they are less evolved and more deeply embedded in the envelope. 

For line observations, the greyscale map represents the intensity integrated across the entire velocity range where emission is detected.  The red and blue contours outline the emission originating from red- and blue-shifted velocities in the range indicated for each panel.  Many velocity ranges were tested for each line in an effort to identify coherent velocity structures in the sources, and those shown represent the clearest structures that could be distinguished.  We refer to the total intensity maps and the velocity structure maps as `Full' and `R/B', respectively.  Velocity ranges and moment zero rms values for both sets of maps are listed in Table \ref{tab:class0i_fluxes}.  We note that the spatial resolution of these observations ($\sim$0.5$\arcsec$; Table \ref{tab:class0i_fluxes}) is not high enough to definitively identify the presence of Keplerian disks \citep[see ][]{Martin-Domenech2019}.  Still, we can identify rotational structures in the compact molecular line emission that are perpendicular to the outflow direction, which we identify as either a disk or the disk-forming region of the rotating inner envelope.  The following sections present the line emission velocity structures identified for each source. 

\subsubsection{Ser-emb 1}
In Ser-emb 1, CO emission clearly traces an outflow structure.  The line emission from C$^{18}$O, H$^{13}$CN, DCN, and HC$^{15}$N show compact structures coincident with the continuum peak.  For C$^{18}$O, there is clearly resolved east-west velocity structure indicative of rotation, which may also be present for HC$^{15}$N though the resolution is too low to confirm.  Interestingly, the extended C$_2$H emission shows rotation in the same east-west direction.  That the compact line emission rotates perpendicular to the outflow is suggestive that it is tracing a disk.  However see \citet{Martin-Domenech2019} for a more detailed analysis of the structure of Ser-emb 1, in which a rotating infalling envelope cannot be ruled out as the origin of velocity structure.  The CO and C$^{18}$O velocity structures are much more coherent in Ser-emb 1 than the other Class 0 sources (Ser-emb 7 and 8), suggesting a smaller envelope contribution and that its structure and evolutionary stage may be more similar to those of a typical Class I source.

\subsubsection{Ser-emb 7}
\label{sec:vels_7}
Ser-emb 7 shows the most complicated structure of the sources in our sample.  No coherent velocity structure could be identified for CO.  For other detected molecular lines (C$^{18}$O, H$^{13}$CN, C$_2$H, and DCN), we identify two molecular emission peaks north and south of the continuum center, which are blue-shifted and red-shifted with respect to one another.  For our analysis we will focus on the the southern molecular emission peak, which is closer to the continuum peak.  While the emission extracted from this region is not necessarily associated with the formation of a disk, it still offers constraints on the chemistry in the inner protostellar core.  The velocities for Ser-emb 7 are shown relative to the southern component velocity center in Figure \ref{fig:serp_vel_summary}, and so all emission in the northern component appears blue-shifted.  C$^{18}$O and C$_2$H do not show any clear velocity structure around the southern component, while H$^{13}$CN and DCN show a velocity gradient along the southwest-northeast axis.  

\subsubsection{Ser-emb 8}
CO in Ser-emb 8 traces a messy outflow structure, and C$^{18}$O shows extended emission with velocity structures roughly similar to CO (i.e., blue-shifted emission in the northeast and red-shifted in the southwest).  H$^{13}$CN, HC$^{15}$N, and DCN all show compact emission with signatures of a southwest-northeast rotation direction.  Though projection effects are difficult to disentangle, the velocity structures are consistent with a small-scale rotation roughly in and out of the plane of the sky, and an outflow direction roughly within the plane of the sky, i.e. perpendicular outflow and disk-like structures.  C$_2$H emission is extended and shows no clear velocity structures.

\subsubsection{Ser-emb 15}
Ser-emb 15 shows the clearest velocity structures of all the sources in our sample.  There is an outflow traced by CO in the north-south direction.  C$^{18}$O, H$^{13}$CN, and DCN show compact emission with east-west rotational structure.  C$_2$H rotates along a similar direction, but with slabs of emission peaking above and below the continuum center.  This is reminiscent of the `hamburger-bun' structure seen towards the edge-on protostellar disk HH-212 \citep{Lee2017, Codella2019}.  Together with the particularly flattened DCN structure, this is quite suggestive that we are viewing an incipient disk edge-on.

\subsubsection{Ser-emb 17}
CO traces a very wide-angle west-east outflow in Ser-emb 17.  Of the compactly emitting molecules, C$^{18}$O and DCN show a clear north-south rotation direction, while no rotational structure can be resolved for H$^{13}$CN or HC$^{15}$N.  C$_2$H shows both compact and extended emission, with no apparent velocity structure.  As in Ser-emb 1, 8, and 15, the compact line emission rotates in a direction perpendicular to the outflow, suggestive of a disk structure.  

\subsubsection{Spectral extraction}
\label{sec:class0_spec}
As seen in Figure \ref{fig:serp_vel_summary}, the structures of these embedded sources can be quite complex.  Our aim is to isolate the molecular component of the protostellar disks or disk-forming regions.  We therefore extract spectra from a region defined by the DCN emitting region: of all the lines covered here, DCN seems to most consistently trace compact disk-like emission.  This is especially apparent in Ser-emb 15, in which DCN traces a flattened, visibly disk-like feature.  For Ser-emb 7, 8, 15, and 17 we adopt a 6$\sigma$ cutoff for the DCN emission mask, and in Ser-emb 1 we choose an 8$\sigma$ cutoff to isolate the more compact central region.  $\sigma$ values correspond to the rms of the DCN `R/B' maps listed in Table \ref{tab:class0i_fluxes}.  For Ser-emb 7 we restrict the DCN mask to exclude the northern molecular emission peak (see Section \ref{sec:vels_7}).  The angular sizes of the resulting masks are 0.40, 0.18, 0.43, 0.46, and 0.37 square arcseconds for Ser-emb 1, 7, 8, 15, and 17, respectively.

\begin{figure*}
	\includegraphics[width=\linewidth]{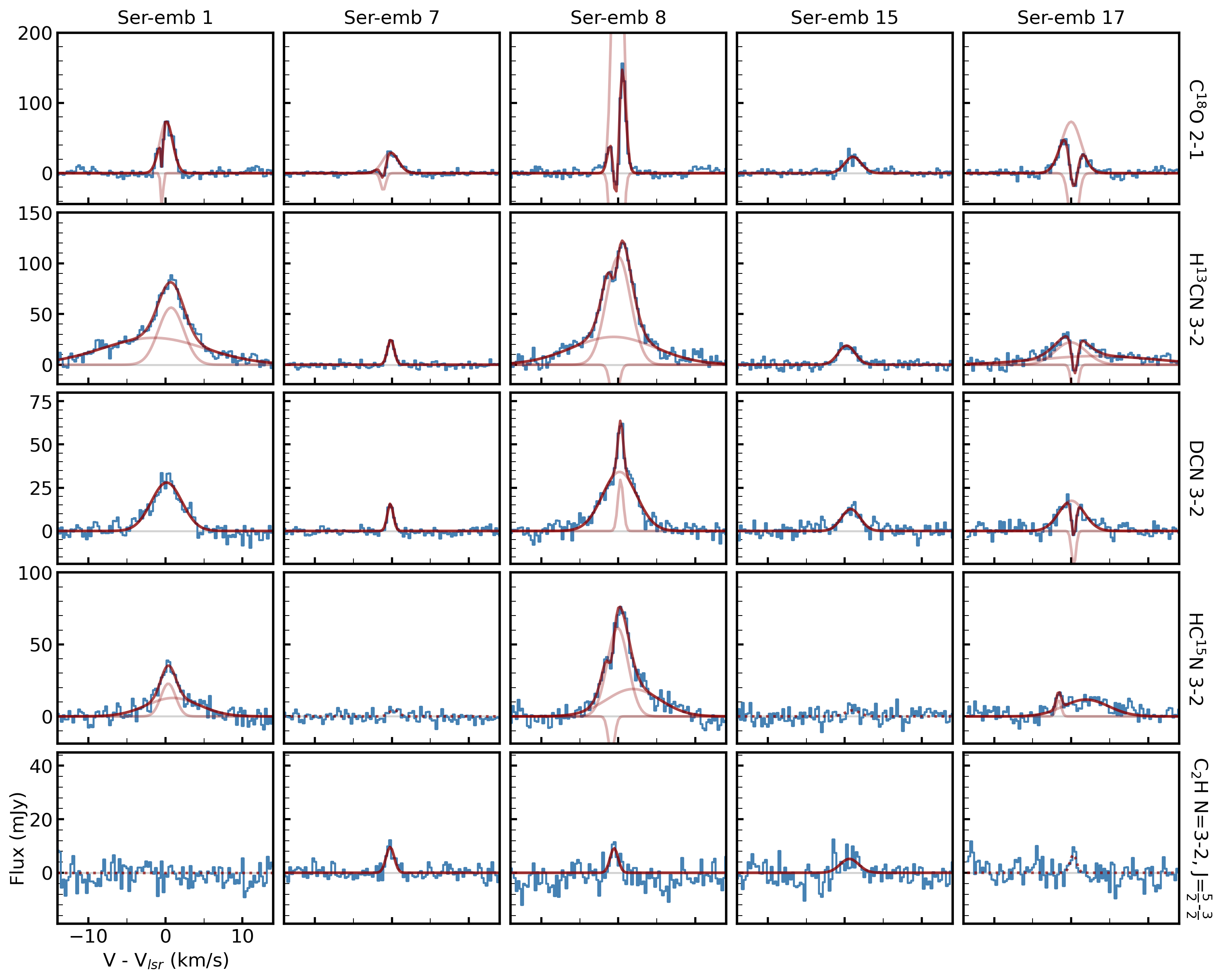}
	\caption{Spectral lines towards the protostellar disk/disk-forming region of the Serpens sources, extracted from within a DCN emission mask.  Blue lines represent the observed spectra.  Dark red lines show the overall fit, with contributions from different velocity components shown as light red lines.  Non-detections ($<$3$\sigma$) are shown as dotted red lines.}
\label{fig:serp_spec}
\end{figure*}

The resulting spectral line profiles are shown in Figure \ref{fig:serp_spec}.  We fit each line as a sum of Gaussians corresponding to different possible velocity components: a `medium' component (Gaussian width constrained to [0.1, 3] km/s) is included for all lines, and other components are added when needed to reproduce the overall profile: `absorption' (Gaussian width [0.1, 1] km/s and amplitude $<$0), `broad' (Gaussian width [2,10] km/s), or `narrow' (Gaussian width [0.1, 0.4]).  With this routine we obtain very good fits to the spectra, as can be seen in Figure \ref{fig:serp_spec}.  We use the MCMC package \texttt{emcee} \citep{Foreman2013} to sample posterior distributions of the Gaussian fit parameters.  By fitting for the absorption feature, we can recover the original flux density prior to self-absorption from the envelope/cloud, though we note that in the case of C$^{18}$O towards Ser-emb 8 and Ser-emb 17 there is some degeneracy between the amplitudes of the emission and absorption features.

We wish to isolate line fluxes originating from the disk-like structures, and excluding emission from the parent cloud or outflows.  We assume that narrow velocity components (emission or absorption, with Gaussian widths $\lesssim$0.5 km/s) arise from the cloud.  We also assume that broad emission components are associated with outflows.  This is supported by the line profiles of DCN (which is spatially most closely associated with compact disk-like structures), which do not include a broad emission component.  Thus, we assume that the flux coming from the disk-like region corresponds to the medium velocity component.  We note that if broad emission does in fact originate from the disk structure, this would increase the fluxes by a factor of $\sim$2--3 for H$^{13}$CN in Ser-emb 1, 8, and 17 and for HC$^{15}$N in Ser-emb 1 and 8.  For non-detections of HC$^{15}$N and C$_2$H, we calculate upper limits adopting the source's line width for H$^{13}$CN and C$^{18}$O, respectively.  The fluxes resulting from this treatment are listed in Table \ref{tab:class0i_fluxes}.  Gaussian fit uncertainties are added in quadrature with a 10\% calibration uncertainty, i.e. the estimated absolute calibration uncertainty for ALMA \citep[e.g.][]{Remijan2020}, and propagated through all subsequent analysis.

\begin{deluxetable*}{lccccccc} 
	\tabletypesize{\footnotesize}
	\tablecaption{Class 0/I line observations \label{tab:class0i_fluxes}}
	\tablecolumns{8} 
	\tablewidth{\textwidth} 
	\tablehead{
		\colhead{Transition}                          &
		\colhead{Beam dim.}                       &
		\colhead{Chan. rms$^a$}                     & 
		\multicolumn{2}{c}{Vel. range$^b$ (km s$^{-1}$)}                          & 
		\multicolumn{2}{c}{Mom. 0 rms$^b$ (mJy beam$^{-1}$ km s$^{-1}$)}                      &
		\colhead{Flux$^c$}                                \\
		\colhead{}                                     & 
		\colhead{($\arcsec \times \arcsec$)}                               &
		\colhead{(mJy beam$^{-1}$)}                 & 
		\colhead{R/B}                                     & 
		\colhead{Full}        & 
		\colhead{R/B}                                     & 
		\colhead{Full}        & 
		\colhead{(mJy km s$^{-1}$)}                    
		  }
\startdata
\hline 
 \multicolumn{8}{c}{Ser-emb 1} \\ 
 \hline 
C$^{18}$O 2-1 & 0.62 $\times$ 0.48 & 5.4 & $|$0.75--1.50$|$ & -1.75--2.00 & 2.8 & 6.1 & 154 $\pm$ 16\\
H$^{13}$CN 3-2 & 0.51 $\times$ 0.43 & 5.0 & $|$1.25--2.00$|$ & -10.75--7.25 & 2.6 & 15.3 & 233 $\pm$ 30\\
DCN 3-2 & 0.62 $\times$ 0.50 & 5.2 & $|$0.75--1.50$|$ & -2.25--2.50 & 2.8 & 6.2 & 139 $\pm$ 15\\
HC$^{15}$N 3-2 & 0.51 $\times$ 0.42 & 5.6 & $|$0.75--1.50$|$ & -1.50--2.50 & 2.9 & 6.2 & 55 $\pm$ 9\\
C$_2$H N=3-2, J=$\frac{5}{2}$-$\frac{3}{2}$ & 0.63 $\times$ 0.51 & 5.0 & $|$0.00--0.50$|$ & -1.00--0.75 & 2.3 & 3.7 & $<$ 21\\
\hline 
 \multicolumn{8}{c}{Ser-emb 7} \\ 
 \hline 
C$^{18}$O 2-1 & 0.61 $\times$ 0.48 & 5.5 & $|$0.00--0.25$|$ & -3.00--3.00 & 1.9 & 7.7 & 74 $\pm$ 9\\
H$^{13}$CN 3-2 & 0.51 $\times$ 0.43 & 5.0 & $|$0.00--1.25$|$ & -2.25--1.00 & 3.3 & 5.1 & 27 $\pm$ 3\\
DCN 3-2 & 0.62 $\times$ 0.51 & 5.2 & $|$0.00--1.25$|$ & -2.25--0.75 & 3.6 & 5.4 & 15 $\pm$ 2\\
HC$^{15}$N 3-2 & 0.51 $\times$ 0.43 & 5.5 & $|$0.00--1.25$|$ & -2.25--0.75 & 3.6 & 5.4 & $<$ 10\\
C$_2$H N=3-2, J=$\frac{5}{2}$-$\frac{3}{2}$ & 0.63 $\times$ 0.51 & 4.9 & $|$0.25--0.50$|$ & -2.50--0.50 & 1.8 & 5.0 & 13 $\pm$ 2\\
\hline 
 \multicolumn{8}{c}{Ser-emb 8} \\ 
 \hline 
C$^{18}$O 2-1 & 0.63 $\times$ 0.51 & 5.3 & $|$0.75--2.00$|$ & -3.00--2.00 & 3.5 & 7.1 & 886 $\pm$ 240\\
H$^{13}$CN 3-2 & 0.52 $\times$ 0.44 & 5.1 & $|$3.50--6.00$|$ & -8.50--5.75 & 4.6 & 12.7 & 426 $\pm$ 50\\
DCN 3-2 & 0.61 $\times$ 0.51 & 5.6 & $|$1.50--6.00$|$ & -3.75--3.75 & 6.4 & 7.9 & 190 $\pm$ 20\\
HC$^{15}$N 3-2 & 0.51 $\times$ 0.43 & 5.8 & $|$1.50--6.00$|$ & -3.50--4.25 & 6.8 & 8.8 & 212 $\pm$ 40\\
C$_2$H N=3-2, J=$\frac{5}{2}$-$\frac{3}{2}$ & 0.63 $\times$ 0.51 & 4.8 & $|$0.00--0.50$|$ & -3.00--1.00 & 2.2 & 5.6 & 12 $\pm$ 4\\
\hline 
 \multicolumn{8}{c}{Ser-emb 15} \\ 
 \hline 
C$^{18}$O 2-1 & 0.63 $\times$ 0.48 & 5.2 & $|$0.50--1.75$|$ & -1.50--2.75 & 3.6 & 6.5 & 62 $\pm$ 7\\
H$^{13}$CN 3-2 & 0.51 $\times$ 0.42 & 4.9 & $|$0.25--2.00$|$ & -1.75--1.50 & 3.8 & 5.0 & 52 $\pm$ 6\\
DCN 3-2 & 0.63 $\times$ 0.50 & 4.9 & $|$0.25--2.00$|$ & -1.25--2.00 & 3.9 & 5.3 & 39 $\pm$ 6\\
HC$^{15}$N 3-2 & 0.51 $\times$ 0.42 & 5.7 & $|$0.25--2.00$|$ & -0.50--1.50 & 4.3 & 4.5 & $<$ 36\\
C$_2$H N=3-2, J=$\frac{5}{2}$-$\frac{3}{2}$ & 0.63 $\times$ 0.51 & 4.9 & $|$0.00--1.00$|$ & -1.50--1.00 & 3.1 & 4.6 & 15 $\pm$ 4\\
\hline 
 \multicolumn{8}{c}{Ser-emb 17} \\ 
 \hline 
C$^{18}$O 2-1 & 0.63 $\times$ 0.50 & 5.3 & $|$1.00--3.00$|$ & -3.00--3.00 & 4.5 & 7.5 & 230 $\pm$ 26\\
H$^{13}$CN 3-2 & 0.52 $\times$ 0.44 & 5.0 & $|$1.00--3.50$|$ & -4.25--6.25 & 4.8 & 10.1 & 112 $\pm$ 19\\
DCN 3-2 & 0.63 $\times$ 0.53 & 5.4 & $|$0.50--3.50$|$ & -2.25--4.50 & 5.2 & 7.5 & 72 $\pm$ 9\\
HC$^{15}$N 3-2 & 0.53 $\times$ 0.44 & 5.5 & $|$0.50--3.50$|$ & -2.00--4.25 & 5.5 & 8.1 & 85 $\pm$ 10\\
C$_2$H N=3-2, J=$\frac{5}{2}$-$\frac{3}{2}$ & 0.63 $\times$ 0.51 & 5.0 & $|$0.00--0.50$|$ & -1.25--1.25 & 2.3 & 4.7 & $<$ 35\\
\enddata
\tablenotetext{}{$^a$ For 0.25 km s$^{-1}$ channels.  $^b$ Velocity ranges and moment zero rms values correspond to two different sets of moment zero maps: `R/B' maps include a subset of channels with emission, chosen to reveal velocity structures, and are shown as red/blue contours in Figure \ref{fig:serp_vel_summary}.  `Full' maps include the full range of channels with emission, shown in greyscale in Figure \ref{fig:serp_vel_summary}.  Fluxes listed in this table correspond to the `Full' velocity range.  $^c$ Extracted within a mask defined by the DCN emitting region (see Section \ref{sec:class0_spec}).  3$\sigma$ upper limits are reported for nondetections.  Uncertainties consist of Gaussian fit uncertainties, added in quadrature with a 10\% calibration uncertainty.}
\end{deluxetable*}

\begin{deluxetable*}{llcccccl} 
	\tabletypesize{\footnotesize}
	\tablecaption{Spectral line data \label{tab:lines}}
	\tablecolumns{8} 
	\tablewidth{\textwidth} 
	\tablehead{
		\colhead{Molecule}                          &
		\colhead{Transition}                               & 
		\colhead{Frequency}                     & 
		\colhead{E$_{up}$}                      & 
		\colhead{log(A$_{ul}$)}                       & 
		\colhead{g$_u$}                       & 
		\colhead{Q(30 K)}                   &
		\colhead{Refs.$^a$}                       \\
		\colhead{}                                     & 
		\colhead{}                               &
		\colhead{(GHz)}                            & 
		\colhead{(K)}                 &
		\colhead{(s$^{-1}$)}                       & 
		\colhead{}						&
		\colhead{}						&
		\colhead{}                 
}                           
\startdata
CO & 2--1 & 230.538 & 16.6 & -6.16 & 5 & 11 & 1 \\ 
$^{13}$CO & 2--1 & 220.399 & 15.9 & -6.22 & 10 & 23 & 2  \\ 
C$^{18}$O & 2--1 & 219.560 & 15.8 & -6.22 & 5 & 12 & 2,3 \\ 
C$_2$H$^b$ &N=3--2, J=$\frac{5}{2}$--$\frac{5}{2}$, F=3--3 & 262.209 & 25.2 & -5.40 & 7 & 59 & 4,5,6 \\ 
&N=3--2, J=$\frac{7}{2}$--$\frac{5}{2}$, F=4--3 \& F=3--2$^c$ & 262.004 \& 262.006 & 25.2 & -4.28 &16 & 59 & 4,5,6 \\
HCN & 3--2 & 265.886 & 25.5 & -3.08 & 21 & 43 & 7 \\ 
H$^{13}$CN & 3--2 & 259.012 & 24.9 & -3.11 & 21 & 44 & 8,9,10 \\ 
DCN & 3--2 & 217.239 & 20.9 & -3.34 & 21 & 53 & 11 \\ 
HC$^{15}$N & 3--2 & 258.157 & 24.8 & -3.12 & 7 & 15 & 8,12\\ 
\enddata
\tablenotetext{}{$^a$ All line parameters are taken from the CDMS catalog \citep{Muller2001, Muller2005} based on data from the following: [1] \citet{Winnewisser1997}, [2] \citet{Klapper2001}, [3] \citet{Winnewisser1985}, [4] \citet{Muller2000}, [5] \citet{Padovani2009}, [6] \citet{Sastry1981}, [7] \citet{Ahrens2002}, [8] \citet{Fuchs2004}, [9] \citet{Cazzoli2005b}, [10] \citet{Maiwald2000}, [11] \citet{Brunken2004}, [12] \citet{Cazzoli2005a} \\
$^b$ The J=$\frac{5}{2}$--$\frac{5}{2}$ transition was observed towards the Class 0/I sources, and the J=$\frac{7}{2}$--$\frac{5}{2}$ complex towards the Class II sources. \\
$^c$ The hyperfine components are blended, so the line parameters refer to the total J=$\frac{7}{2}$--$\frac{5}{2}$ line.  The total degeneracy is found by adding the degeneracies of each hyperfine component, and the total Einstein coefficient is found from the degeneracy-weighted sum of Einstein coefficients of each hyperfine component.}
\end{deluxetable*}

\subsection{Class II sources}
\label{sec:classii_dets}

\begin{figure*}
\centering
	\includegraphics[width=\linewidth]{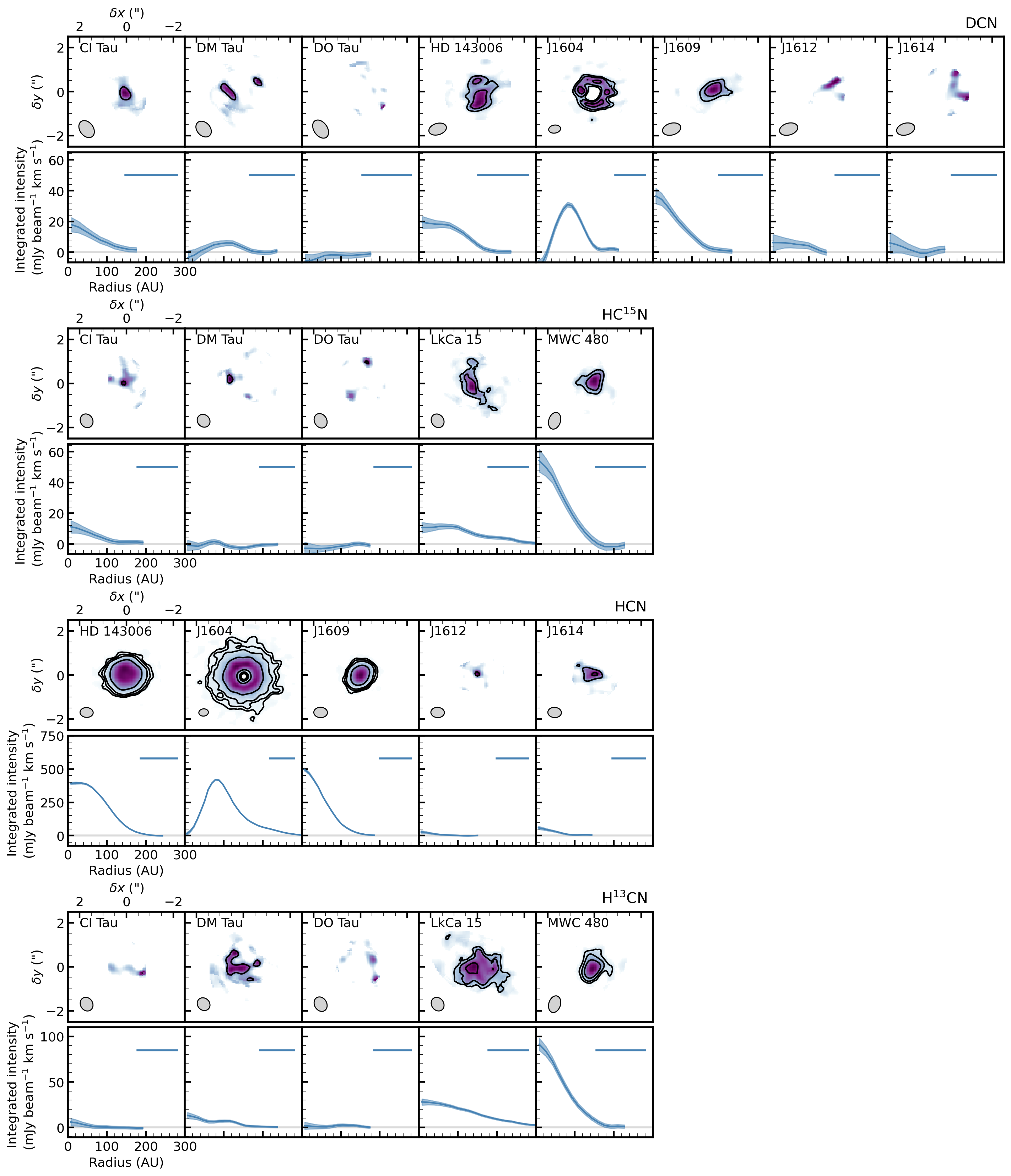}
	\caption{HCN isotopologues in the Class II disks.  Top sub-panels: moment zero maps extracted using Keplerian masking.  Contours represent 3, 5, 8, 20, 50, 100$\times$ the median rms across the map.  The rms values for each panel can be found in Table \ref{tab:classii_fluxes}.  Emission at a level below 1$\times$rms is not shown, and color scales are normalized to each individual image.  The restoring beams are shown in the lower left of each image.  Bottom sub-panels: deprojected, azimuthally averaged intensity profiles.  Shaded regions show the 1$\sigma$ uncertainties at each radial distance, calculated as $\sigma_{mom0}$/$\sqrt{N}$, where $\sigma_{mom0}$ is the moment zero rms and $N$ is the number of independent measurements (i.e. beams) per annulus.  Horizontal bars in the top right of each panel represent the restoring beam major axis.}
\label{fig:classii_d15}
\end{figure*}

Figure \ref{fig:classii_d15} shows the moment zero maps for the HCN isotopologues observed towards the Class II disks, along with deprojected and azimuthally averaged intensity profiles.  $^{13}$CO moment zero maps for J1609, J1612, and J1614 can be found in Appendix \ref{sec:app_lineobs}.  Keplerian masking was used to generate the moment zero maps, assuming the star properties listed in Table \ref{tab:sourcedat}, and disk inclinations and position angles listed in Appendix \ref{sec:app_lineobs} (Table \ref{tab:disk_geom}).  Mask outer radii are chosen to encompass the observed emission.  A full description of the masking technique can be found in \citet{Bergner2018} and \citet{Pegues2020}.  Deprojected radial profiles were generated using the disk inclinations and position angles in Table \ref{tab:disk_geom}.

Fluxes are calculated by summing the emission within the moment zero map.  We also apply the same Keplerian mask to 200 synthetic image cubes of randomly drawn emission-free channels neighboring the target line.  The integrated flux uncertainty is estimated from the standard deviation in the integrated fluxes of these emission-free moment zero maps, added in quadrature with a 10\% calibration uncertainty.  The moment zero rms varies across the map because each pixel represents the sum of a different number of channels when Keplerian masking is used.  We therefore calculate the standard deviation in intensity within each pixel across the 200 emission-free maps, and adopt the median pixel rms as the moment zero map rms.  Line observation results for the Class II sources are listed in Appendix \ref{sec:app_lineobs} (Table \ref{tab:classii_fluxes}).

We consider a line to be detected if the measured flux is $>$3$\sigma$ and if emission $>$3$\sigma$ is present in the moment zero map.  Based on these detection criteria, DCN is detected towards CI Tau, HD 143006, J1604, and J1609; HC$^{15}$N towards LkCa 15 and MWC 480; HCN towards HD 143006, J1604, J1609, and J1614; and H$^{13}$CN towards DM Tau, LkCa 15, MWC 480.

The HCN and H$^{13}$CN lines presented here were imaged with different parameters from \citet{Bergner2019b} in order to be consistent with the DCN and HC$^{15}$N images, thus facilitating our analysis of isotopic ratios.  The main difference in imaging is in the choice of robust parameter.  Here, weak lines are imaged with a robust parameter of 2.0 instead of 1.0, which enabled a detection of H$^{13}$CN towards DM Tau and of HCN towards J1614.  We also imaged HCN towards J1604 with a robust of 0.0 since the emission is very bright, allowing us to better resolve the emission morphology in the inner disk.  We also note that H$^{13}$CN and HC$^{15}$N 3--2 were previously observed towards LkCa 15 and MWC 480 with a similar sensitivity and spatial resolution in \citet{Guzman2017}.  We obtain H$^{13}$CN fluxes a factor of $\sim$1.3 and 1.7 higher compared to \citet{Guzman2017} for MWC 480 and LkCa 15, respectively, possibly due to our Keplerian masking technique which reduces the incorporation of emission-free regions into our moment zero images.  We find a similar HC$^{15}$N flux in MWC 480 but a slightly lower flux in LkCa 15 (by a factor of $\sim$0.8) compared to \citet{Guzman2017}.  This is reflective of the uncertainty inherent in imaging weak molecular line data; our uncertainties are probably under-estimated for lines close to the detection threshold.
  
In disks where both DCN and HCN are detected, their morphologies are qualitatively similar: HD 143006 shows a  plateau in intensity towards the disk center, J1604 has a strong emission depletion towards the center, and J1609 is centrally peaked.  The inner gap in J1604 is wider in DCN than in HCN, and the intensity is negative within $\sim$30 AU.  This negative emission is not seen for channels without line emission, and could be caused by over-subtraction of the continuum where it is coupled to the line emission.  That the DCN intensity drops below zero while HCN does not may indicate that it is emitting closer to the midplane.  HC$^{15}$N and H$^{13}$CN exhibit shallow, extended emission profiles in LkCa 15 and a centrally peaked morphology in MWC 480.  HC$^{15}$N emission in LkCa 15 is asymmetrical and lacks emission in the northeast region of the disk, whereas H$^{13}$CN emission is more axisymmetric. 

\section{HCN, C$_2$H, and C$^{18}$O column densities \& abundances}
\label{sec:cd_abunds}
\subsection{Class 0/I column densities}
With the integrated flux densities $\int S_\nu d\nu$ from Table \ref{tab:class0i_fluxes}, we estimate total molecular column densities $N_T$ from:

\begin{equation}
N_T = \frac{4\pi \int S_\nu d\nu}{A_{ul}\Omega h c g_u} e^{E_{u}/T_r} Q(T_r), 
\label{eq:Nt}
\end{equation}
where $A_{ul}$ is the Einstein coefficient for the transition, $\Omega$ is the angular size of the emitting region, $g_u$ is the upper state degeneracy, $E_u$ is the upper state energy, $T_r$ is the rotational temperature, and $Q(T_r)$ is the temperature-dependent molecular partition function.  All spectral line parameters can be found in Table \ref{tab:lines}.  For $\Omega$, we adopt the angular size of the masks used for spectral extraction (listed in Section \ref{sec:class0_spec}).  The rotational temperatures are not known and cannot be determined from these data alone, so we calculate column densities adopting rotational temperatures of 20, 30, and 50 K.  This is informed by the lukewarm temperature profile derived for the embedded disk L1527 \citep[$\sim$20--50 K midplane temperatures; ][]{vantHoff2018}, though the effect of a much warmer (100 K) rotational temperature is discussed in Section \ref{sec:disc}.

The resulting column densities are shown in Table \ref{tab:serp_cd}.  We find that the choice of rotational temperature has only modest effects on the column density.  For subsequent analysis, we adopt the 30 K case as our fiducial value, and incorporate the variation in column density resulting from a 20 K or 50 K rotational temperature into the uncertainties.

\begin{deluxetable}{lr} 
	\tabletypesize{\footnotesize}
	\tablecaption{Class 0/I column densities \label{tab:serp_cd}}
	\tablecolumns{2} 
	\tablewidth{\textwidth} 
	\tablehead{
		\colhead{Source}                          &
		\colhead{$N_{T, 30K}$ [$\Delta$ $_{20K}$, $\Delta$ $_{50K}$] (cm$^{-2}$)}                                                                        
		  }
\startdata
\hline 
 \multicolumn{2}{c}{C$^{18}$O} \\ 
 \hline 
Ser-emb 1 & 6.8 $\pm$ 1.7 $\times$10$^{15}$ [-0.8, +2.3]\\
Ser-emb 7 & 6.0 $\pm$ 1.5 $\times$10$^{15}$ [-0.7, +2.0]\\
Ser-emb 8 & 3.6 $\pm$ 1.3 $\times$10$^{16}$ [-0.4, +1.2]\\
Ser-emb 15 & 2.5 $\pm$ 0.7 $\times$10$^{15}$ [-0.3, +0.8]\\
Ser-emb 17 & 1.1 $\pm$ 0.3 $\times$10$^{16}$ [-0.1, +0.4]\\
\hline 
 \multicolumn{2}{c}{H$^{13}$CN} \\ 
 \hline 
Ser-emb 1 & 9.8 $\pm$ 1.6 $\times$10$^{12}$ [+0.2, +1.8]\\
Ser-emb 7 & 2.1 $\pm$ 0.3 $\times$10$^{12}$ [+0.0, +0.4]\\
Ser-emb 8 & 1.6 $\pm$ 0.3 $\times$10$^{13}$ [+0.0, +0.3]\\
Ser-emb 15 & 2.0 $\pm$ 0.3 $\times$10$^{12}$ [+0.0, +0.4]\\
Ser-emb 17 & 4.9 $\pm$ 1.0 $\times$10$^{12}$ [+0.1, +0.9]\\
\hline 
 \multicolumn{2}{c}{C$_2$H} \\ 
 \hline 
Ser-emb 1 & $<$ 7.0 $\times$10$^{14}$ [+0.2, +1.3]\\
Ser-emb 7 & 7.7 $\pm$ 1.5 $\times$10$^{14}$ [+0.2, +1.4]\\
Ser-emb 8 & 3.5 $\pm$ 1.2 $\times$10$^{14}$ [+0.1, +0.6]\\
Ser-emb 15 & 4.6 $\pm$ 1.3 $\times$10$^{14}$ [+0.1, +0.8]\\
Ser-emb 17 & $<$ 1.2 $\times$10$^{15}$ [+0.0, +0.2]\\
\hline 
 \multicolumn{2}{c}{DCN} \\ 
 \hline 
Ser-emb 1 & 1.0 $\pm$ 0.2 $\times$10$^{13}$ [-0.0, +0.3]\\
Ser-emb 7 & 2.1 $\pm$ 0.4 $\times$10$^{12}$ [-0.1, +0.5]\\
Ser-emb 8 & 1.3 $\pm$ 0.2 $\times$10$^{13}$ [-0.1, +0.3]\\
Ser-emb 15 & 2.6 $\pm$ 0.5 $\times$10$^{12}$ [-0.1, +0.7]\\
Ser-emb 17 & 5.6 $\pm$ 1.1 $\times$10$^{12}$ [-0.3, +1.4]\\
\hline 
 \multicolumn{2}{c}{HC$^{15}$N} \\ 
 \hline 
Ser-emb 1 & 2.3 $\pm$ 0.5 $\times$10$^{12}$ [+0.0, +0.4]\\
Ser-emb 7 & $<$ 7.5 $\times$10$^{11}$ [+0.1, +1.4]\\
Ser-emb 8 & 8.3 $\pm$ 1.8 $\times$10$^{12}$ [+0.2, +1.5]\\
Ser-emb 15 & $<$ 1.4 $\times$10$^{12}$ [+0.0, +0.3]\\
Ser-emb 17 & 3.8 $\pm$ 0.6 $\times$10$^{12}$ [+0.1, +0.7]\\
\enddata
\tablenotetext{}{Column densities and uncertainties are calculated assuming a 30 K rotational temperature.  Numbers listed in brackets indicate the change to the listed column density for a 20 K or 50 K rotational temperature, respectively (in the same order of magnitude as the 30 K value).}
\end{deluxetable}

\subsection{Class II column densities}
\label{sec:classii_molcd}
Column density estimates for C$_2$H, HCN, and C$^{18}$O in the Class II sources are derived using the observations presented in \citet{Bergner2019b}, with a few exceptions.  First, for CI Tau, DM Tau, DO Tau, LkCa 15, and MWC 480, we do not have ALMA observations of HCN 3--2, so we previously used HCN fluxes measured with the SMA for these disks.  However, for the comparison presented here, we opt to solve for the HCN column density using H$^{13}$CN as a proxy since HCN suffers from optical depth effects \citep{Bergner2019b}.  Second, because C$^{18}$O 2--1 is not detected towards J1609, J1612, or J1614, we previously used CO to estimate C$^{18}$O fluxes.  However, here we use $^{13}$CO instead, as CO is likely optically thick in these sources.  These two changes allow for an improved comparison of the chemistry across our sample.  

To estimate disk-averaged column densities, integrated fluxes are extracted from moment zero maps using an elliptical mask.  The mask is defined by the maximum radius containing emission at a 4$\sigma$ level, projected along the same position angle and inclination used to generate the Keplerian mask (Table \ref{tab:disk_geom}).  This treatment allows us to determine the disk-integrated flux and the emission region $\Omega$ self-consistently, while avoiding severe beam dilution; note though that these fluxes differ slightly from those integrated across the entire moment zero map, as in Section \ref{sec:classii_dets}.  For non-detected lines, we derive upper limits using the continuum major axis (fit at the 3--5$\sigma$ level) as the mask radius.

Equation \ref{eq:Nt} was used along with the spectral line parameters listed in Table \ref{tab:lines} to estimate column densities from disk-integrated fluxes.  As for the Class 0/I sources, we solve for column densities assuming rotational temperatures of 20, 30, and 50 K.  The integrated fluxes, emitting regions $\Omega$, and resulting column densities are listed in Table \ref{tab:classii_cd} in Appendix \ref{sec:app_cd}.  As in the case of the Serpens sources, the column densities are not very sensitive to the choice in rotational temperature. 

As demonstrated in \citet{Bergner2019b}, the C$_2$H and HCN lines covered here are generally optically thick in Class II disks.  We therefore adjust the Class II HCN and C$_2$H column densities following \citet{Goldsmith1999}:
\begin{equation}
N_{T} = N_{T, \mathrm{thick}}\times\frac{\tau}{1-e^{-\tau}}.
\label{eq:tau}
\end{equation}
This correction excludes HCN column densities derived from H$^{13}$CN, as these are already likely to be optically thin.  

For the disks fit in \citet{Bergner2019b}, the average disk-averaged optical depth for the C$_2$H N=3-2, J=$\frac{7}{2}$--$\frac{5}{2}$ and HCN J=3--2 lines are 2.1$\pm$0.4 and 5.8$\pm$1.8, respectively.  Uncertainties represent the standard deviations across the disk sample, and are propagated throughout all analyses involving $\tau$-corrected column densities.  Note that these values represent the total line optical depth of all hyperfine components, $\tau_{\rm{total}}$.  In principle, the flux from each hyperfine component should be corrected individually for optical depth, however in the disk-integrated spectra used in this work, the hyperfine components are blended.  Because of this, the optical depth corrections are approximate.  For C$_2$H, the F=4--3 and F=3--2 hyperfine components have similar relative intensities (0.572 and 0.428, respectively), and the estimated $\tau_{\rm{total}}$ is not too high.  We therefore assume that the opacity of the blended hyperfine lines can be described by the weighted opacity of the individual components, leading to $\tau_{\rm{blend}}$ = 0.51$\times \tau_{\rm{total}}$.  For the C$_2$H N=3--2, J=$\frac{7}{2}$--$\frac{5}{2}$ $\tau_{\rm{total}}$ of 2.1$\pm$0.4, this results in a $\tau_{\rm{blend}}$ of 1.1$\pm$0.2 for use in Equation \ref{eq:tau}.  In the case of HCN, most opacity (92.6\%) in the J=3--2 transition is contained in the overlapping F=2--1, 3--2, 4--3, and 2--3 components at 265.886 GHz.  These components are sufficiently overlapping that they may be treated as a single line.  We assume that the  $\tau_{\rm{total}}$ of 5.8$\pm$1.8 can be used directly in Equation \ref{eq:tau}, though this may over-correct the contribution from the satellite lines (F=3--3 and F=2--2) which are less optically thick than the main complex. 

\subsection{HCN, C$_2$H, and C$^{18}$O correlations across evolutionary stages}
\label{sec:hcn_c2h_c18o}
We now combine the Class 0/I and Class II column density results to explore correlations between HCN, C$_2$H, and C$^{18}$O across all stages.  The size scales that the column densities are extracted from correspond to $\sim$150 AU for the Class 0/I sources, and $\sim$150--600 AU for the Class II sources.  For sources with H$^{13}$CN observations (i.e., all Class 0/I and some Class II sources), we convert from H$^{13}$CN to HCN column densities using the local ISM $^{12}$C/$^{13}$C ratio of 68$\pm$15 \citep{Milam2005}.  We note that $^{13}$C fractionation is not expected to be important for HCN in protostellar or disk environments \citep[e.g.][]{Roberts2002, Hily-Blant2019}.  For J1609, J1612, and J1614, C$^{18}$O column densities are found from $^{13}$CO assuming the local ISM $^{16}$O/$^{18}$O ratio of 560$\pm$25 \citep{Wilson1994} and the same $^{12}$C/$^{13}$C ratio as before.  

We also estimate disk-averaged H$_2$ column densities for our source sample to enable a comparison of molecular abundances.  Following \citet{Hildebrand1983} and \citet{Andrews2005}, we convert the 260 GHz continuum flux to a disk mass by:
\begin{equation}
M_d = \frac{d^2 F_\nu}{\kappa_v \zeta B_\nu(T)},
\end{equation}
where $M_d$ is the disk mass, $d$ is the distance, $\kappa_v$ is the dust opacity, $\zeta$ is the dust-to-gas ratio, and $B_\nu(T)$ is the Planck function given a dust temperature T. We adopt a $\kappa_v$ of 2.62 cm$^2$ g$^{-1}$ following the power-law scaling of \citet{Beckwith1990} with a power-law index $\beta$=1.  We assume a dust-to-gas ratio of 1:100 and a uniform dust temperature of 20 K \citep{Andrews2005}.  We acknowledge that the dust opacity, dust-to-gas ratio, dust size distribution, and dust temperature may vary across our source sample, and further insight into the physical structures of these sources is needed for a more robust abundance comparison.  The effects of these uncertainties are discussed further in Section \ref{sec:caveats}.

We convert the disk mass $M_d$ to the total number of H$_2$ molecules $\mathcal{N}_{H2}$ assuming a typical gas molecular weight of 2.37$\times$ the mass of atomic hydrogen.  Finally, we convert to an H$_2$ column density by dividing $\mathcal{N}_{H2}$ by the area of the mask used to extract the continuum flux.  For the Class 0/I sources this is the same DCN emission mask used for the spectral line extraction.  Note that for Ser-emb 7, this does not coincide with the continuum peak position.  For the Class II sources, we integrate the continuum flux within an ellipse fit to the 260 GHz continuum at a 3--5$\sigma$ threshold, depending on the continuum brightness.  While the Class II continuum emission is generally more compact than the molecular line emission (typically by a factor of $\sim$2--5 in angular area; see Tables \ref{tab:classii_cd} and \ref{tab:cont_h2}), we assume that the H$_2$ column densities derived in this way are representative of the disk average.  Continuum fluxes, source areas $\Omega$, and H$_2$ columns are listed in Table \ref{tab:cont_h2} in Appendix \ref{sec:app_cd}.

\begin{figure*}
\centering
	\includegraphics[width=\linewidth]{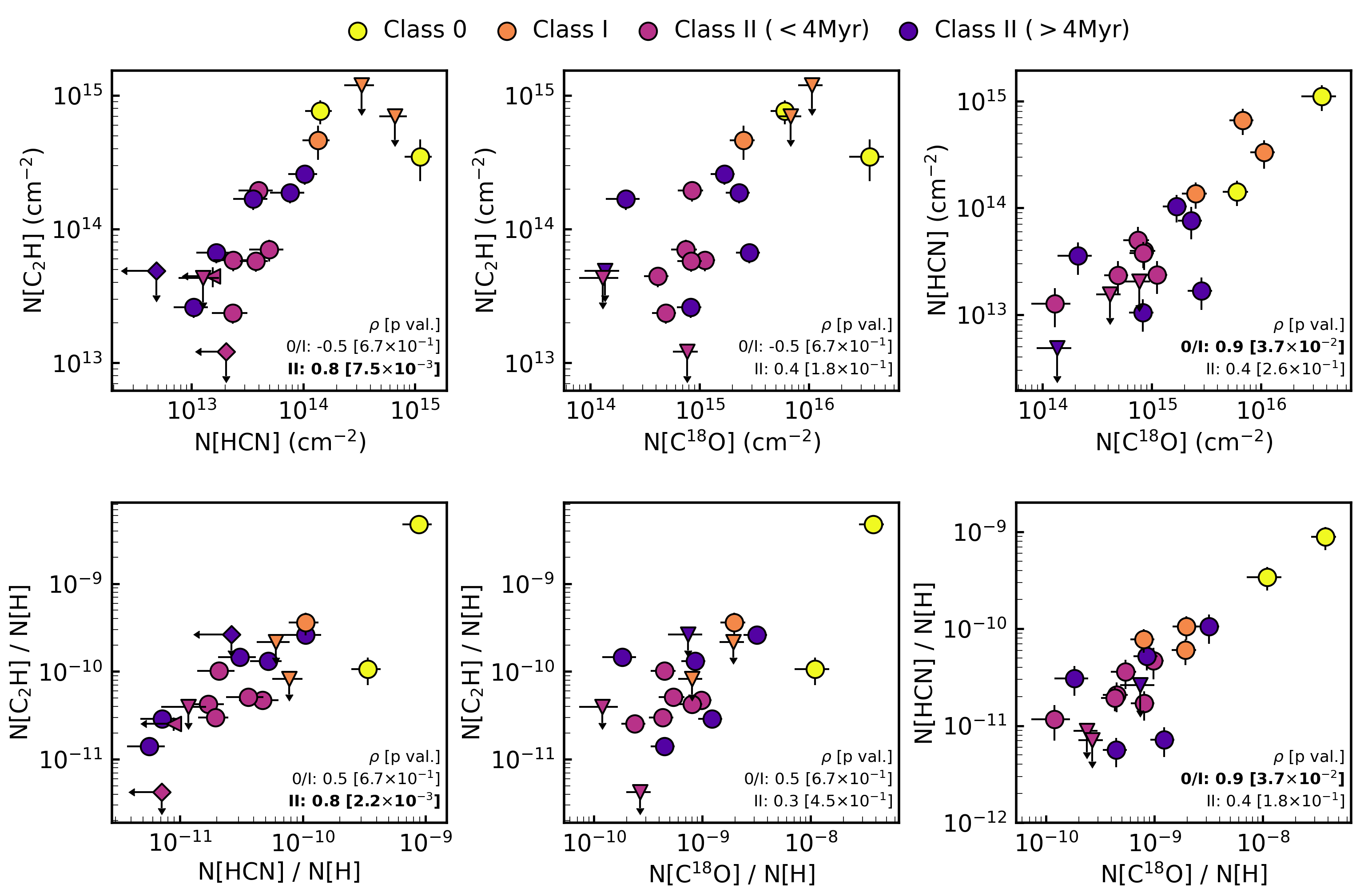}
	\caption{Correlation plots for the disk-averaged C$_2$H, HCN, and C$^{18}$O column densities (top) and abundances with respect to H (bottom).  Yellow, orange, pink, and purple points represent Class 0, Class I, young Class II ($<$4 Myr), and old Class II ($>$4 Myr) sources, respectively.  Triangle and diamond markers indicate sources with one or both upper limits for each molecule pair, respectively.  Spearman $\rho$ correlation coefficients are shown for the Class 0/I and the Class II sources; statistically significant values are bolded.  Ser-emb 1 is included in the Class I category based on its CO isotopologue morphologies (Section \ref{sec:class0_structures}) and its chemical similarity to the other Class I sources.}
\label{fig:cno} 
\end{figure*}

Figure \ref{fig:cno} shows comparisons of the C$_2$H, HCN, and C$^{18}$O column densities (top) and column density ratios with respect to H (bottom), which we will refer to as abundances.  For each panel, we also include the Spearman $\rho$ correlation coefficient calculated for the Class 0/I sources and the Class II sources.  This measures the degree to which the column densities/abundances of each molecule pair are monotonically related.  Possible $\rho$ values range from -1 (strongly anti-correlated) to 1 (strongly correlated), with 0 indicating no correlation.    We use the \texttt{scipy.stats.spearmanr} module to calculate Spearman $\rho$ coefficients and their p values.  We consider a p value $<$5\% to be statistically significant.  Note that this treatment excludes upper limits.  

The observed correlations are similar when considering column densities or abundances, and our discussion will mainly focus on the abundances.  Statistically significant positive correlations are seen between the HCN and C$_2$H abundances in the Class II sources, and between the HCN and C$^{18}$O abundances in the Class 0/I sources.  In all other cases, abundance correlations are weakly positive and statistically insignificant.

There are no clear abundance jumps/discontinuities between the Class 0/I and Class II sources.  Within the protostellar stage, abundances are typically higher in the Class 0 sources than the Class I sources.  Within the Class II sources, there are no clear trends between $<$4 Myr and $>$4 Myr sources.  Chemical implications of these trends are discussed in Section \ref{sec:disc_cno}.

It is important to note that our sample contains just two Class 0 and three Class I sources, and a larger sample size is needed for a more robust assessment of correlation and scatter within these evolutionary stages.  Also, better constraints for sources with non-detections are needed to improve this analysis.  In particular, the C$_2$H transition observed towards the Class 0/I sources is over an order of magnitude weaker than the transition observed towards the Class II sources, and the upper limits for non-detections are not very constraining.

\subsection{C$^{18}$O abundances vs. age}
\label{sec:c18o_age}
\begin{figure}
\centering
	\includegraphics[width=\linewidth]{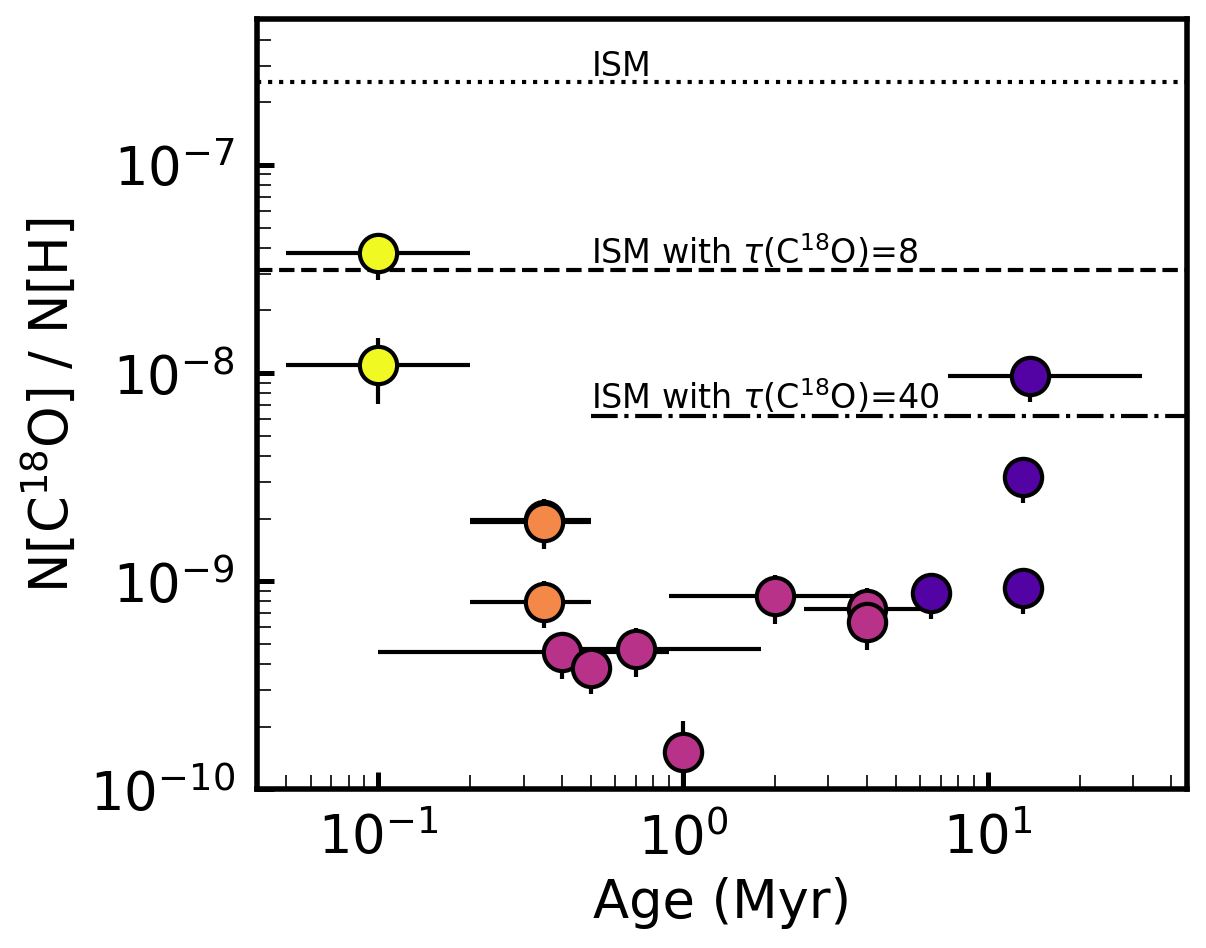}
	\caption{Age dependence of C$^{18}$O/H abundances.  Colors are the same as in Figure \ref{fig:cno}.  Note that these values are measured within the inner beam to minimize freeze-out contributions at larger radii, and thus may differ from the disk-averaged values in Figure \ref{fig:cno}.  The C$^{18}$O/H abundance expected for the ISM is shown as a dotted line, and the apparent abundance given an optical depth of 8 and 40 as a dashed line and dot-dash line, respectively. }
\label{fig:co_age} 
\end{figure}

With these observations, we can also evaluate how the C$^{18}$O abundance changes with evolutionary stage.  Here, for the Class II disks, we re-derived column densities and abundances using C$^{18}$O and continuum fluxes extracted from a circle defined by the C$^{18}$O beam major axis.  By focusing on the inner-most disk regions accessible at this resolution, we aim to reduce the impact of CO freeze-out on our observations.  We also exclude J1609, J1612, and J1614 from this analysis, as their C$^{18}$O abundances are estimated from $^{13}$CO which may be optically thick.  Class II source ages and uncertainties are taken from Table \ref{tab:sourcedat}.  For the Class 0 and Class I sources, we adopt ages of 0.1$\pm$0.1 and 0.35$\pm$0.15 Myr, respectively.  This is based on the estimated Class 0 and Class I lifetimes of 0.2 and 0.5 Myr, respectively \citep{Enoch2009, Evans2009}.  

Figure \ref{fig:co_age} shows the resulting C$^{18}$O abundances as a function of the estimated source ages.  We emphasize that the C$^{18}$O optical depth is a major uncertainty in assessing both the absolute C$^{18}$O/H abundances and the relative trends between evolutionary stages.  Still, we can use constraints from the literature to estimate how much our observations are affected.  C$^{18}$O emission has been found to be optically thick in several embedded disks \citep{vantHoff2018, Zhang2020}.  The C$^{18}$O 2--1 optical depths in the embedded disks in \citet{Zhang2020} are likely around $\sim$4--8, based on the reported C$^{18}$O and $^{13}$C$^{18}$O fluxes and assuming a $^{12}$C/$^{13}$C ratio around 70.  We therefore expect that moderate optical depth effects may affect the protostellar sources in our sample.  C$^{18}$O has also been found to be optically thick within the CO snow lines of Class II disks, with opacities ranging from $\sim$8 in TW Hya \citep{Zhang2017} to several tens in HD 163296 \citep{Booth2019}.  It is therefore probable that C$^{18}$O optical depth effects range from moderate to severe for the Class II disks in our sample. 

In light of this, it is likely not appropriate to compare our measured C$^{18}$O/H abundances to the ISM abundance of $\sim$2.5$\times$10$^{-7}$ (assuming CO/H = 1.4$\times$10$^{-4}$ and $^{16}$O/$^{18}$O = 560).  We therefore include in Figure \ref{fig:co_age} the apparent ISM abundance given an optical depth of 8 and 40 as more realistic benchmarks.  Based on the previously mentioned literature measurements, we estimate that the $\tau$=8 case is the most appropriate comparison for the Class 0/I sources.  The Class II disks likely span a range of optical depths between the $\tau$=8 and $\tau$=40 cases.  As a sanity check, we compare the depletion factors inferred from our C$^{18}$O/H abundances to the depletion factors found from detailed modeling in \citet{Zhang2019}.  We recover similar CO depletion factors for HD 163296 and DM Tau in the $\tau$=40 scenario (i.e., $\sim$1-10$\times$ depleted), and for IM Lup in the $\tau$=8 scenario ($\sim$10-100$\times$ depleted).  This further supports that this is a reasonable range of optical depths to assume for the Class II disks in our sample.  Still, we stress the need for observations of a rarer CO isotopologue to better constrain the magnitude of optical depth effects in our observations.  

 As seen in Figure \ref{fig:co_age}, the Class 0 C$^{18}$O/H abundances in our sample are consistent with ISM levels if C$^{18}$O is moderately optically thick ($\tau$=8).  On the other hand, even assuming $\tau$=8, the Class I C$^{18}$O/H abundances are depleted in CO relative to ISM levels by about an order of magnitude.  The Class II sources are likely depleted in CO by factors of a few up to 100, depending on the optical depths in individual sources.  We cannot distinguish if there are any trends in CO depletion with age within the Class II stage given the optical depth uncertainties.  However, even in the $\tau$=40 case we see that most disks exhibit around an order of magnitude in CO depletion relative to ISM levels.  The exceptions, V4046 Sgr and J1604, show abundances that are comparable to ISM levels if the C$^{18}$O optical depths are large.

It is important to consider possible effects of CO freeze-out on our derived abundances.  Although we extract fluxes from the inner beam of the Class II disks to minimize freeze-out effects, in some sources the midplane CO snow line radius may be smaller than the beam.  Still, while freeze-out can lower the gas-phase CO abundance in the dense midplane region, detailed models show that freeze-out around the midplane cannot explain the low CO abundances in the emissive surface layers which our observations are sensitive to \citep{Favre2013, Zhang2019}.  Moreover, for the disks around the Herbig Ae stars HD 163296 and MWC 480, which have a warmer temperature structure and midplane snow lines likely beyond the radius of our extraction mask, we measure C$^{18}$O/H abundances comparable to those of other Class II sources.  Thus, CO freeze-out cannot fully explain the low CO abundances observed in our sample.  For the embedded disks, we expect that the temperature structures are warmer \citep{vantHoff2018}, and CO freeze-out should not impact the inner protostellar core.  Still, it is important to recognize that episodic accretion events may play an important role in setting the CO freeze-out behavior in protostars \citep{Jorgensen2015}.  Improved constraints on the protostellar physical structures and evolutionary histories are needed to better understand whether accretion variations are likely to impact our conclusions.

\subsection{Abundance caveats}
\label{sec:caveats}
There are several additional uncertainties that may affect the abundances we derive for all molecules.  Optically thick continuum emission could be problematic for recovering accurate molecular line fluxes \citep{Cleeves2016, Harsono2018}.  Additionally, our assumption of a constant dust opacity for all sources may introduce bias into this comparison, as $\kappa$ depends on the grain size distribution and could therefore vary systematically with source age.  However, it is not clear that there is a more appropriate assumption, as dust opacities are not well constrained for any source age.  Extrapolating $\kappa_{1.3\rm{mm}}$ = 1.0 cm$^2$g$^{-1}$ derived for dense protostellar cores in \citet{Ossenkopf1994} (assuming $\beta$=1) yields a $\kappa_{1.1\rm{mm}}$ about 2.5$\times$ smaller than the value that we used.  If the true opacity is smaller than our adopted value, this would result in an under-estimated H$_2$ column and over-estimated molecular abundances for the Class 0/I sources.  Similarly, if the dust-to-gas ratio increases with evolutionary stage, this would result in an over-estimated H$_2$ column and in turn under-estimated molecular abundances in the older sources. 

Another uncertainty in deriving abundances is the temperature of the gas and dust emitting in the protostellar sources.  For instance, gas in the inner protostellar regions may be warmer than 30 K.  If we instead adopt an excitation temperature of 100 K in the Class 0/I sources, this increases the column densities of all molecules by a factor of $\sim$2.  Additionally, because the disk cools and becomes more optically thick from the Class 0 to Class I stage, the adopted 20 K characteristic dust temperature may result in mass over-estimates for the younger protostellar sources or under-estimates for the older protostellar sources \citep{Jorgensen2009, Dunham2014b}.  Adopted dust temperatures of 15 K or 30 K instead of 20 K results in $\sim$40--50\% variations in the derived dust masses (and therefore H$_2$ columns).

\section{HCN isotopologue ratios}
\label{sec:hcn}
\subsection{Class 0/I sources}
\label{sec:class0I_hcn}
As discussed in Section \ref{sec:intro}, isotopic fractionation occurs under specific physical conditions, and fractionation patterns are often used to infer the conditions in which molecules formed.  We aim to explore deuterium and $^{15}$N fractionation using our observations of the HCN isotopologues.  Because the major isotopologue HCN is not covered for the Class 0/I sources, we convert from H$^{13}$CN to HCN fluxes assuming the local interstellar $^{12}$C/$^{13}$C ratio of 68 $\pm$ 15 \citep{Milam2005}.  As mentioned in Section \ref{sec:hcn_c2h_c18o}, $^{13}$C fractionation is not expected to be important in protostellar or disk environments \citep[e.g.][]{Roberts2002, Hily-Blant2019}.  HCN/DCN and HCN/HC$^{15}$N ratios are then found from the ratio of column densities determined in Section \ref{sec:class0_spec}.  

The HCN isotopologue ratios are listed in Table \ref{tab:hcn_ratios} for an assumed rotational temperature of 30 K.  HCN/HC$^{15}$N ratios range from 88 in Ser-emb 17 to 287 in Ser-emb 1.  For comparison, the $^{14}$N/$^{15}$N ratio in the local ISM is 450$\pm$22, similar to the value of 441$\pm$5 measured in the Solar wind \citep{Wilson1994, Marty2011}.  Thus, we see some degree of $^{15}$N enhancement in HCN for all of the Serpens sources.  HCN/DCN ratios ratios are comparatively more consistent across the Serpens sources, ranging only from 52 in Ser-emb 15 to 87 in Ser-emb 8.  Relative to the local ISM H/D ratio of 4.3$\pm$0.4$\times$10$^4$ and the protosolar value of 5.0$\pm$0.9$\times$10$^4$ \citep{Linsky2006, Geiss2003}, these sources show significant deuterium fractionation. 

As noted previously, the choice of rotational temperature has a small impact on our results.  This is particularly true for column density ratios: changing the rotational temperature from 30 to 50 K or 30 to 20 K results in HCN/DCN variations around 6\% and HCN/HC$^{15}$N variations around 1\%.  This is small compared to other sources of uncertainty (i.e. the $^{12}$C/$^{13}$C ratio, Gaussian line fitting, and telescope calibration).  In addition to these known sources of uncertainty, we emphasize that this treatment assumes that the HCN isotopologues emit co-spatially.  Higher spatial resolution and multi-line observations are needed to confirm if all isotopologues are indeed characterized by the same emitting area and rotational temperature.

\begin{deluxetable*}{lcccc} 
	\tabletypesize{\footnotesize}
	\tablecaption{HCN isotopologue flux \& column density ratios \label{tab:hcn_ratios}}
	\tablecolumns{5} 
	\tablewidth{\textwidth} 
	\tablehead{
		\colhead{Source}                          &
		\colhead{Major isotopologue}    &
		\colhead{$F$(Major/Minor)$^a$} & 
		\multicolumn{2}{c}{$N_T$(HCN/Minor) $^{b,c}$} \\
		\colhead{} & 
		\colhead{} & 
		\colhead{} & 
		\colhead{No $\tau$ correction} &
		\colhead{$\tau$(HCN)=5.8$\pm$1.8}		                                                      
		  }
\startdata
\hline 
 \multicolumn{5}{c}{HC$^{15}$N} \\ 
 \hline 
Ser-emb 1 & H$^{13}$CN & 4.3 [0.9] & 287 [97]\\
Ser-emb 7 & H$^{13}$CN & $>$2.8 & $>$189\\
Ser-emb 8 & H$^{13}$CN & 2.0 [0.4] & 135 [47]\\
Ser-emb 15 & H$^{13}$CN & $>$1.4 & $>$97\\
Ser-emb 17 & H$^{13}$CN & 1.3 [0.3] & 88 [30]\\
\hline 
DM Tau & H$^{13}$CN  & $>$1.7 & $>$115 & \\
LkCa 15 & H$^{13}$CN  & 1.5 [0.3] & 100 [31] & \\
MWC 480 & H$^{13}$CN  & 1.6 [0.3] & 106 [29] & \\
\hline 
 \multicolumn{5}{c}{DCN} \\ 
 \hline 
Ser-emb 1 & H$^{13}$CN & 1.7 [0.3] & 65 [21]\\
Ser-emb 7 & H$^{13}$CN & 1.8 [0.3] & 68 [23]\\
Ser-emb 8 & H$^{13}$CN & 2.2 [0.4] & 87 [28]\\
Ser-emb 15 & H$^{13}$CN & 1.3 [0.3] & 52 [18]\\
Ser-emb 17 & H$^{13}$CN & 1.5 [0.3] & 60 [21]\\
\hline 
CI Tau & H$^{13}$CN  & $<$0.9 & $<$33 & \\
DM Tau & H$^{13}$CN  & $>$1.8 & $>$70 & \\
HD 143006 & HCN  & 24.5 [2.8] & 13 [2] & 75 [25]\\
J1604 & HCN  & 18.4 [0.6] & 10 [1] & 56 [18]\\
J1609 & HCN  & 17.8 [2.6] & 9 [2] & 54 [19]\\
J1614 & HCN  & $>$5.4 & $>$3 & $>$17\\
\enddata
\tablenotetext{}{
$^a$ The ratio of the integrated line intensities of the major and minor HCN isotopologues. \\
$^b$ Column density ratios are calculated assuming a rotational temperature of 30 K. \\
$^c$ When H$^{13}$CN is the major isotopologue, we use the $^{12}$C/$^{13}$C ratio to convert from H$^{13}$CN to HCN, and assume the emission is optically thin.  When HCN is the major isotoplogue, we show the column density ratio assuming either optically thin emission, or adjusting for an estimated optical depth.  See text for details.} 
\end{deluxetable*}

\subsection{Class II sources}
\label{sec:classII_hcn}
We now estimate the HCN/DCN and HCN/HC$^{15}$N ratios in the Class II disks with detections.  In some disks H$^{13}$CN was observed instead of HCN, so like for the Class 0/I disks we convert from H$^{13}$CN to HCN fluxes assuming a $^{12}$C/$^{13}$C ratio of 68 $\pm$ 15 \citep{Milam2005}.  When performing this analysis, it is important that the fluxes are extracted from the same spatial region.  The CASA task \texttt{imsmooth} is used to smooth the HCN and H$^{13}$CN images to the same resolution as the DCN images.  We then use the same mask to extract fluxes of each major and minor isotopologue pair (i.e., HCN or H$^{13}$CN, and HC$^{15}$N or DCN).  We define the extraction mask as pixels with $>$3$\sigma$ emission of both the major and minor isotopologue in order to avoid diluting the intensity of the minor isotopologue.  Note that the fluxes used in this analysis differ from the disk-integrated values presented previously.  We solve for the major:minor isotopologue column density ratio according to:

\begin{equation}
\frac{N_{T}(a)}{N_{T}(b)} = \frac{\int S_\nu d\nu (a)}{\int S_\nu d\nu (b)} \frac{Q(T_r) (a)}{Q(T_r) (b)} \frac{e^{E_u (a)/T_r}}{e^{E_u (b)/T_r}} \frac{A_{ul} (b) g_u (b)}{A_{ul} (a) g_u (a)},
\end{equation}
where all parameters are the same as described for Equation \ref{eq:Nt}.  As before, we adopt a rotational temperature of 30 K.  Variations in the isotopologue ratios derived using 20 K or 50 K rotational temperatures are incorporated into the uncertainties.  As described in Section \ref{sec:classii_molcd}, HCN is likely optically thick in the Class II disks.  We therefore apply an opacity correction assuming $\tau$=5.8$\pm$1.8 (see Section \ref{sec:classii_molcd}) to the HCN fluxes used in this analysis.

Table \ref{tab:hcn_ratios} lists the derived HCN isotopologue column density ratios.  The HCN/HC$^{15}$N ratio could be measured for two sources in the sample.  In LkCa 15 and MWC 480, the HCN/HC$^{15}$N ratios of 100 and 106, respectively, are low compared to the local ISM and protosolar values around 450 \citep{Wilson1994, Marty2011}, and within the range measured for the Serpens sources (Section \ref{sec:class0I_hcn}).  The ratios we derive are similar to those found in \citet{Guzman2017}.  In DM Tau, H$^{13}$CN is detected but HC$^{15}$N is not detected, so we place a lower limit on HCN/HC$^{15}$N of 115. Neither HC$^{15}$N nor H$^{13}$CN was detected towards CI Tau or DO Tau, and so we cannot place any useful limits on the HCN/HC$^{15}$N ratio in those sources.

The HCN/DCN ratio could be measured for three sources: HD 143006, J1604, and J1609.  For these sources, the HCN/DCN ratio is $\sim$10 assuming HCN is optically thin, or 54--75 for an HCN opacity of 5.8.  We can also place limits on the HCN/DCN ratio for CI Tau of $<$33, for DM Tau of $>$70, and for J1614 of $>$17 (assuming $\tau_{\rm{HCN}}$=5.8).  Note that CI Tau is detected in DCN but not H$^{13}$CN (Section \ref{sec:classii_dets}), resulting in an HCN/DCN upper limit.  In DO Tau and J1612, neither DCN nor the major isotopologue (HCN or H$^{13}$CN) is detected, so no limits can be placed. 

For comparison, the Class II HCN/DCN ratios measured from (optically thin) H$^{13}$CN/DCN flux ratios in \citet{Huang2017} range from $\sim$14-194 assuming a 30 K excitation temperature.  Both the opacity-uncorrected and corrected values that we find are consistent with this range.  Regardless of the optical depth, the ratios we measure are very low compared to the local ISM and presolar values around 5$\times$10$^4$ \citep{Linsky2006, Geiss2003}.  Compared to the Class 0/I HCN/DCN ratios around 70, the ratios in the Class II sources are either low in the optically thin case or comparable in the optically thick case.  

\subsection{HCN fractionation across evolutionary stages}
\label{sec:hcn_frac}

\begin{figure*}
\centering
	\includegraphics[width=\linewidth]{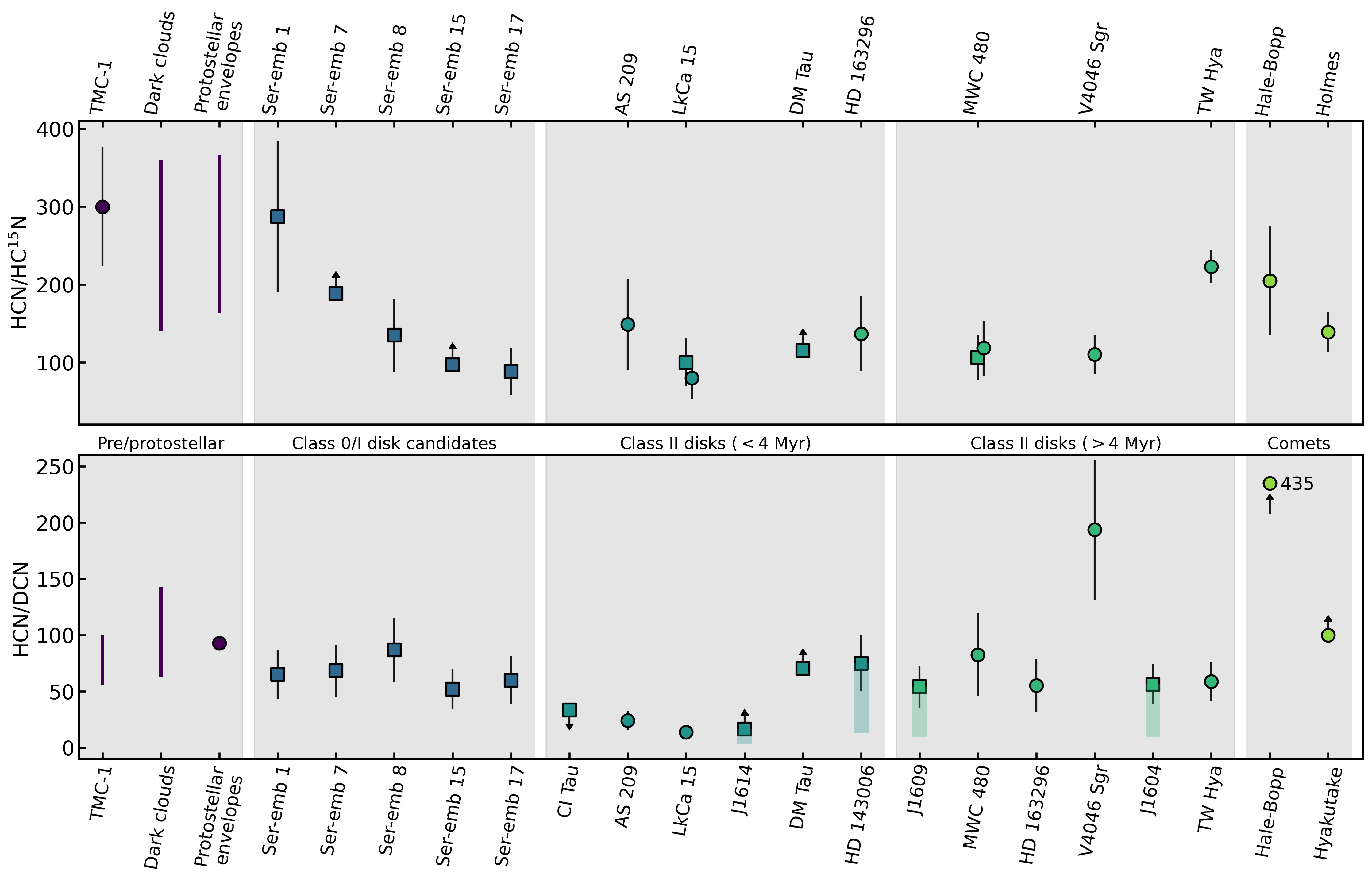}
	\caption{HCN isotopologue ratios during the star formation sequence (see text for citations of literature measurements).  For the pre- and protostellar sources, vertical bars represent the range of values measured towards a source or group of sources.  Shaded bars show the range of HCN/DCN ratios assuming an HCN optical depth up to 5.8 (see Section \ref{sec:classII_hcn}).}
\label{fig:isotopologues}
\end{figure*}

Our observations cover multiple HCN isotopologues in both the Class 0/I and Class II sources, enabling a comparison of HCN/HC$^{15}$N and HCN/DCN ratios throughout the disk lifetime.  Figure \ref{fig:isotopologues} shows the resulting column density ratios derived in Sections \ref{sec:class0I_hcn} and \ref{sec:classII_hcn}.  For comparison, we also show literature measurements of other evolutionary stages: HCN/HC$^{15}$N in the pre- and proto-stellar sources TMC-1 \citep{Ikeda2002}, the dark clouds L183 and L1544 \citep{Hily-Blant2013}, and the protostellar envelopes of IRAS 16293A, R CrA IRS7B, and OMC-3 MMS6 \citep{Wampfler2014}; the Class II disks AS 209, LkCa 15, MWC 480, HD 163296, V4046 Sgr, \citep{Guzman2017} and TW Hya \citep{Hily-Blant2019}; and the comets Hale-Bopp and Holmes \citep{Bockelee-Morvan2008}.  For HCN/DCN we include measurements in the pre- and proto-stellar sources TMC-1 \citep{Turner2001}, a sample of infrared dark clouds \citep{Feng2019}, and a sample of protostellar envelopes \citep{Jorgensen2004}; the Class II disks AS 209, LkCa 15, MWC 480, HD 163296, V4046 Sgr, \citep{Huang2017} and TW Hya \citep{Qi2008}; and the comets Hale-Bopp \citep{Meier1998} and Hyakutake \citep{Bockelee-Morvan2008}.   For the Class II measurements taken from \citet{Guzman2017} and \citet{Huang2017}, we re-derive column density ratios assuming a 30 K rotational temperature to enable a consistent comparison with our results.  

The HCN/HC$^{15}$N ratios in the Class 0/I sources range from $\sim$90--290, and decrease with evolutionary stage as traced by bolometric temperature.  Most Class II disks have an HCN/HC$^{15}$N ratio between $\sim$80 and 150.  The exception, TW Hya, has a disk-averaged ratio above 200, though the ratio in the inner disk is lower \citep[121 $\pm$ 11; ][]{Hily-Blant2019}.  The HCN/HC$^{15}$N ratios in the Class II disks are comparable to the Class 0/I sources Ser-emb 8 and 17 (and the lower limit in Ser-emb 15), and low compared to the other pre- and protostellar sources.  Cometary HCN/HC$^{15}$N ratios are consistent with (if on the high end) of the values measured in Class II disks.

Compared to the HCN/HC$^{15}$N ratio, the HCN/DCN ratio is quite flat across our sample of Class 0/I disk candidates, ranging from just 52--87.  These ratios are comparable to measurements in pre- and protostellar sources.  In the Class II sources, the new measurements we derive have large uncertainties due to potential optical depth effects (Section \ref{sec:classII_hcn}).  Even so, it seems that the HCN/DCN ratio is slightly lower in young Class II sources compared to the protostellar disk sources.  The HCN/DCN ratio again increases slightly in $>$3 Myr Class II sources, beginning with DM Tau.  Cometary HCN/DCN ratios are high compared to most Class II disks.  The exception is V4046 Sgr, whose anomalously high HCN/DCN ratio is consistent with the upper limit measured towards Hale-Bopp.

\section{Discussion}
\label{sec:disc}

Our sample spanning Class 0/I protostellar disk candidates and Class II disks allows us to explore evolutionary trends in disk chemistry.  In Section \ref{sec:hcn_c2h_c18o} we looked for evolutionary relationships between C$_2$H, HCN, and C$^{18}$O chemistries, and in Section \ref{sec:hcn_frac} for patterns in $^{14}$N/$^{15}$N and D/H fractionation of HCN.  We now discuss the chemical implications of our analysis.

\subsection{Evolution of C$^{18}$O abundances}
\label{sec:disc_c18o}

In Section \ref{sec:c18o_age} we presented evolutionary patterns in C$^{18}$O/H abundances.  While there are a number of important uncertainties in our derived abundances, namely the C$^{18}$O optical depths, we can still identify several trends from the existing data.  We emphasize that more rigorous abundance retrievals and/or observations of rarer CO isotopologues are needed to confirm our findings.  

The Class 0 C$^{18}$O/H abundances are not distinguishable from ISM levels with our data, assuming that C$^{18}$O optical depths up to a factor of $\sim$10 are possible in these sources.  It is also important to note that the disk cannot always be isolated in the Class 0 sources: the morphology of C$^{18}$O emission in the Class 0 sources (Figure \ref{fig:serp_vel_summary}) is more suggestive of envelope emission than protostellar core/disk emission.  Observations of a rarer isotopologue would help to isolate emission from the inner disk-forming region \citep[e.g.][]{Zhang2020}.

The C$^{18}$O/H abundances in the Class I sources are difficult to reconcile with an ISM abundance, even assuming moderate C$^{18}$O optical depths.  It is likely that these sources are depleted in CO by about an order of magnitude compared to ISM levels.  We note that \citet{Zhang2020} recently found evidence for ISM-like CO abundances in three protostellar disks, which may signify that a range of depletion timescales and outcomes are possible.  Indeed, chemical models by \citet{Drozdovskaya2014} demonstrate that differences in infall physics can strongly influence the chemical processing that occurs in embedded disks.  The sources in \citet{Zhang2020} are among the most massive protostellar disks known, and their higher CO abundances may reflect a different physical/chemical environment compared to the Serpens sources.  Still, the results from this work and \citet{Zhang2020} both indicate that CO depletion is a rapid process taking place on a $\sim$0.5--1 Myr timescale.

The decrease in C$^{18}$O/H abundances from the Class 0 to Class I sources is either an artifact of envelope material clouding our view of the Class 0 core/disk regions, or reflects an active depletion mechanism.  The drop in CO abundances also coincides with an increase in $^{15}$N fractionation of HCN, which is likely due to enhanced photochemistry during envelope clearing (see Section \ref{sec:disc_hcn}).  It is possible that envelope clearing also facilitates CO depletion, e.g. through increased exposure to X-rays which drive CO destruction by reactions with He$^+$ \citep[e.g.][]{Favre2013}.  We also note that single-dish studies show that CO abundances on envelope scales increase from Class 0 to Class I objects \citep{Jorgensen2002, Jorgensen2005}, converse to what we see here.  This implies that the chemistry in the inner protostellar regions is distinct from the envelope, and highlights the importance of understanding chemical evolution during infall.  

The Class II sources have a wide range of probable C$^{18}$O optical depths, from factors of a few to several tens \citep{Zhang2017, Booth2019}, making it difficult to interpret depletion trends with age within the Class II stage.  Still, even with a conservative estimate of C$^{18}$O optical depth, $\tau$=40, the majority of disks exhibit C$^{18}$O/H abundances about an order of magnitude below ISM levels.  The similar depletion levels in the Class II sources compared to the Class I sources suggests that the low CO abundances seen widely in Class II disks \citep[e.g.][]{Dutrey2003, Favre2013, Cleeves2016, Zhang2017} are set very early in the disk lifetime.  Indeed, \citet{Cleeves2016} suggest depletion in an earlier evolutionary stage as a possible explanation for CO depletion in the young disk IM Lup \citep[$\sim$0.5 Myr; ][]{Andrews2018}. 

While most disks have apparent C$^{18}$O/H abundances around 10$^{-10}$--10$^{-9}$, two sources, V4046 Sgr and J1604, have notably higher abundances closer to 10$^{-8}$.  Interestingly, both are old ($\sim$13 Myr) transition disks.  The high C$^{18}$O abundances in these disks could be explained by a distinct chemistry taking place at the heavily irradiated inner disk edge, or could be an artifact of different dust properties skewing the derived gas masses in (old) transition disks compared to full disks.  

In summary, our results are consistent with a scenario in which CO is depleted relative to ISM levels on a $\sim$0.5--1 Myr timescale.  We do not see evidence for significantly higher depletion levels in the Class II stage compared to the Class I stage.  This suggests that depletion levels are largely set in the embedded stages, and further depletion during the Class II stage is not especially efficient.  An important consideration if CO depletion takes place early is that models of physical and chemical mechanisms for CO depletion in disks generally begin with ISM CO abundances and assume physical properties of Class II disks \citep[e.g][]{Reboussin2015, Krijt2018, Schwarz2018, Eistrup2018}.  Our results underscore the need to test whether these (or other) mechanisms could deplete CO in the context of protostellar rather than protoplanetary disk physics and chemistry.  

\subsection{C$_2$H, HCN, and C$^{18}$O chemistries}
\label{sec:disc_cno}
We now assemble evidence from the C$_2$H, HCN, and C$^{18}$O emission morphologies (Figure \ref{fig:serp_vel_summary}) and abundance correlations (Figure \ref{fig:cno}) to gain insight into the chemical and physical factors driving their formation.  In particular, we wish to explore how the chemistries of the three molecules relate to one another.  As described in Section \ref{sec:intro}, models predict that CO will be removed from the disk atmosphere over its lifetime, as some combination of physical and chemical processes sequester volatiles in the midplane \citep[e.g.][]{Meijerink2009, Krijt2016, Schwarz2018}.  In an increasingly O-poor (i.e., high C/O) gas in the disk atmosphere, HCN and C$_2$H formation are predicted to be enhanced \citep{Du2015}.  The expectation would therefore be that CO abundances decrease and HCN and C$_2$H abundances increase as the disk evolves.  While our results do not fit this evolutionary scenario, the underlying chemistry (i.e. efficient C$_2$H and HCN formation in O-depleted gas) can still explain most of our results.

\subsubsection{Class 0/I sources}
A positive correlation is seen between HCN and C$^{18}$O abundances in the protostellar sources.  These molecules are also spatially coincident, with both generally tracing rotating material in the inner protostellar core.  It is important to note that our sample includes just five Class 0/I sources, and a larger sample size is needed to confirm the observed HCN and C$^{18}$O correlation.  Based on the existing data, however, we speculate that the observed positive correlation is related to sublimation of ices from infalling material.  An ice contribution is supported by that the Serpens sources with hot corino emission \citep[Ser-emb 1, 8, and 17; ][]{Bergner2019c} have systematically higher column densities of all HCN isotopologues than the non-hot corino sources (Table \ref{tab:serp_cd}).  It is also otherwise difficult to explain why a tight abundance correlation (and spatial coincidence) would be observed in the protostellar sources but not the Class II sources.  The correlation is likely driven primarily by CO sublimation in the protostellar core, which releases carbon into the gas and facilitates gas-phase formation of C-bearing molecules like HCN.  The similar binding energies of N$_2$ and CO \citep[e.g.][]{Fayolle2016} means that nitrogen should also be abundant in the gas phase in similar regions.  A gas-phase HCN formation channel seems necessary to explain the HC$^{15}$N fractionation patterns seen in the protostellar sources (see Section \ref{sec:disc_hcn}), though direct sublimation of HCN ice may also contribute to its gas-phase abundance.  We also note that HCN emission is more compact than C$^{18}$O in the Class 0 sources.  This could be an excitation effect, or could imply that some additional ingredient beyond gas-phase C and N is facilitating its formation in the disk-like region of the protostar (e.g. moderate radiation exposure or higher densities).

C$_2$H emits in a distinct region compared to HCN and C$^{18}$O in the protostellar sources (Figure \ref{fig:serp_vel_summary}).  The morphologies in Ser-emb 1, 8, and 15 are suggestive of either an outflow cavity or a proto-disk atmosphere, consistent with observations of small hydrocarbons towards other embedded sources \citep[e.g.][]{Oya2014, Artur2019}.  In both cases we expect low shielding from radiation in these regions, supporting that C$_2$H formation is favored by an intense irradiation field \citep{Bergin2016, Aikawa1999}.  Highly irradiated environments may also have high C/O ratios due to the destruction of carbon grains or PAHs \citep[e.g.][]{Visser2007, Anderson2017}, further enhancing C$_2$H formation.
 
\subsubsection{Class II sources}
HCN and C$_2$H show a strong positive abundance correlation in the Class II sources, indicating a common physical/chemical driver.  We do not observe an anti-correlation between C$^{18}$O with C$_2$H or HCN, indicating that factors other than the C/O ratio are influencing the chemistry.  However, this does not mean that CO depletion is unimportant for C$_2$H and HCN formation.  The majority of Class II sources have estimated C$^{18}$O/H abundances that are low compared to the ISM value (Section \ref{sec:disc_c18o}).  The lack of a clear trend between C$^{18}$O with C$_2$H and HCN may simply reflect that the chemistry in these disks is already operating in a high C/O gas.  Indeed, \citet{Cleeves2018} show that C$_2$H and HCN formation become insensitive to changes in the C/O ratio once the value is above $\sim$1.8.  The bright emission of C$_2$H in most Class II disks may itself imply a high C/O ratio in the disk atmosphere: chemical models predict that efficient hydrocarbon formation is favored by C/O ratios $\gtrsim$1 \citep{Bergin2016, Schwarz2019, Miotello2019}.  It is important to note that that achieving a C/O ratio $>$1 requires CO destruction (not simply sequestration), with O being partitioned in e.g. H$_2$O and C remaining in the gas \citep{Bergin2016}.

Thus, if a sufficiently high C/O ratio is present in all disks, then HCN and C$_2$H formation may be insensitive to variations in the C/O ratio, and other factors will become important in shaping the abundance relationships between molecules.  For instance, all three molecules should depend on the availability of gas-phase carbon, which will decrease as CO is removed.  The weak (statistically insignificant) positive correlations between C$_2$H and HCN with C$^{18}$O could in part reflect this common dependency.  Radiation exposure is also predicted to be an essential ingredient for C$_2$H and HCN formation \citep[e.g.][]{Aikawa1999}, and could contribute to the strong correlation between the two molecules. 
 
It is important to note that in some disks, the morphologies of C$_2$H and HCN are quite distinct \citep[see especially DM Tau, HD 163296, and V4046 Sgr in Figure 2 of ][]{Bergner2019b}, indicating that, while the disk-averaged abundances show a clear relationship, there are still important differences in their formation chemistries.  While this work is focused on comparing bulk disk properties, further insight into the relationship between chemical families will be gained by high-resolution observations that compare chemical sub-structures.

\subsection{Evolution of HCN fractionation}
\label{sec:disc_hcn}

As shown in Figure \ref{fig:isotopologues}, the protostellar sources show a decreasing HCN/HC$^{15}$N ratio with evolutionary stage as traced by bolometric temperature.  This indicates active HC$^{15}$N fractionation in the protostellar stage.  This trend is readily explained by photodissociation-driven fractionation turning on as the envelope clears and the disk is increasingly exposed to UV from the protostar.  If temperature-driven fractionation were responsible, we would expect to see a similar trend in the evolution of DCN fractionation.  We note that a photochemical fractionation scenario requires gas-phase HCN formation via N$_2$ dissociation, implying that HCN observed in the protostellar cores is at least in part formed in situ in the gas.  Photochemical fractionation is also thought to be important in Class II disks based on radially resolved HCN/HC$^{15}$N measurements \citep{Guzman2017, Hily-Blant2019}.  The agreement between HCN/HC$^{15}$N ratios in the most evolved protostellar sources and the Class II sources supports that the same mechanism is at play.  Still, our inferences are based on disk-averaged properties, and higher-resolution observations that reveal radial variations in the HCN/HC$^{15}$N ratio would help to confirm whether and how photochemical fractionation takes place in these sources.

Unlike HC$^{15}$N, deuterium fractionation in the protostellar stage is either inefficient or proceeds with comparable efficiency across all Class 0--Class I sources, resulting in flat HCN/DCN ratios.  This likely reflects that deuterium fractionation is typically temperature-driven rather than photodissociation-driven, and thus should not exhibit strong trends with envelope clearing.  With the data available to date, we see a tentative trend in which HCN/DCN ratios decrease slightly from Class I to young Class II disks, followed by a modest increase in HCN/DCN ratios in $>$3 Myr Class II disks (with the exception of the anomalously high increase in V4046 Sgr).  These trends may trace evolution of the physical structures of Class II disks, however a spatially resolved analysis is needed to better understand the dependence of D fractionation patterns on disk physical properties.

\section{Conclusions}
We present ALMA observations of C$^{18}$O, C$_2$H, and HCN (and isotopologues DCN and HC$^{15}$N) towards a sample of protostellar (Class 0/I) disk candidates and protoplanetary (Class II) disks.  We aim to isolate emission associated with the protostellar core/disk-forming region in the Class 0/I sources, enabling a view of chemical evolution throughout the disk lifetime.  We focus our analysis on the relationship between C$^{18}$O, C$_2$H, and HCN, as well as isotopic fractionation of HC$^{15}$N and DCN.  We conclude the following:

\begin{itemize}[leftmargin=*]
\item Coherent velocity structures in molecular line emission are identified on small scales towards the continuum peak of all Class 0/I sources except for Ser-emb 7, which displays a complicated physical structure.  Generally the rotation axis of the small-scale line emission is perpendicular to the outflow orientation, consistent with a disk or disk-forming structure.  DCN 3--2 emission seems to reliably trace the compact disk-like structures of the embedded sources, especially notable in Ser-emb 15.

\item We estimate molecular column densities and abundances for the disk or disk-forming region of all Class 0--Class II sources.  Accounting for probable C$^{18}$O optical depth effects, the C$^{18}$O/H abundances are likely an order of magnitude lower than ISM levels in the Class I sources, and factors of a few to two orders of magnitude lower in the Class II sources.  Our results are consistent with a rapid CO depletion taking place on a timescale of 0.5--1 Myr, and suggest that the low CO abundances seen widely in Class II disks are set very early in the disk lifetime.  

\item C$_2$H and HCN abundances are positively correlated in the Class II stage, reflecting a common physical and/or chemical driver.  We do not observe an anti-correlation in the abundances of C$^{18}$O with HCN or with C$_2$H.  This may reflect that CO is already quite depleted by the Class I stage: given sufficiently high C/O ratios, further CO depletion does not map to enhanced C$_2$H and HCN formation.  Instead, CO depletion may actually hinder HCN and C$_2$H production by limiting the supply of gas-phase carbon.  

\item HCN and C$^{18}$O show a strong positive abundance correlation in the Class 0/I sources.  This is likely driven by ice sublimation from infalling material, with gas-phase HCN formation enhanced by the release of CO (and N$_2$) in the warm protostellar core.  This scenario is supported by that HCN, DCN, and HC$^{15}$N column densities are systematically higher in the Class 0/I sources that also host hot corino emission, which signals efficient ice desorption.  

\item C$_2$H emission traces outflow cavity or proto-disk atmosphere structures in the protostellar sources.  This supports that C$_2$H formation is favored by intense irradiation in PDR-like transition zones.  

\item HCN/HC$^{15}$N levels range from $\sim$90--290 in the Class 0/I sources and decrease with evolutionary stage (as traced by $T_{bol}$).  This can be explained by an increase in photodissociation-driven fractionation as the envelope clears.  The Class II sources show similar HCN/HC$^{15}$N ratios as found in the older protostellar sources ($\sim$80--150), perhaps reflecting a continuing photochemical fractionation throughout the Class II lifetime.

\item HCN/DCN ratios are quite consistent across the protostellar sources ($\sim$50--90), indicating that gas-phase D fractionation is either inefficient or is consistently efficient during the protostellar stage.  HCN/DCN ratios drop slightly between Class I and young Class II sources, with a tentative increase again in $>$3 Myr disks.  This could trace disk physical evolution in the Class II stage, though a spatially resolved analysis is needed to better understand the D fractionation mechanism.
\end{itemize}

From this survey, we see that the embedded stage represents a distinct chemical regime from the Class II stage in several of ways.  Shielding of radiation by the envelope inhibits gas-phase $^{15}$N fractionation pathways, and also means that C$_2$H does not form co-spatially with the dust disk until the Class II stage.  Additionally, ice sublimation from infalling material may provide important ingredients for gas-phase chemistry in the protostellar core.  On the other hand, CO appears depleted early in the disk lifetime, and depletion levels may not vary significantly between embedded disks and Class II disks.  The role of such effects in setting the volatile compositions of planet-forming material is an important question going forward.

There are a number of avenues that will further our understanding of disk chemical evolution.  We emphasize the need for larger sample sizes to better disentangle the effects of disk physical properties on the chemistry through more robust statistical analysis.  Multi-line observations would also provide improved constraints on line optical depths and excitation temperatures, allowing for more accurate column density estimates and offering insight into where vertically the emission originates.  Lastly, local chemical variations are seen even with the moderate spatial resolution in this program, e.g. the distinctive C$_2$H morphologies compared to HCN and C$^{18}$O.  Future high-resolution studies of chemical sub-structures should offer a powerful and complementary tool for understanding how volatile chemistry depends on local physical and chemical disk properties.  

\acknowledgments 
We are grateful to the anonymous referee for feedback on this manuscript.  This paper makes use of ALMA data, project codes 2015.1.00964.S and 2016.1.00627.S.  ALMA is a partnership of ESO (representing its member states), NSF (USA), and NINS (Japan), together with NRC (Canada) and NSC and ASIAA (Taiwan), in cooperation with the Republic of Chile. The Joint ALMA Observatory is operated by ESO, AUI/NRAO, and NAOJ. The National Radio Astronomy Observatory is a facility of the National Science Foundation operated under cooperative agreement by Associated Universities, Inc.  

J.B.B. acknowledges support from NASA through the NASA Hubble Fellowship grant \#HST-HF2-51429.001-A awarded by the Space Telescope Science Institute, which is operated by the Association of Universities for Research in Astronomy, Incorporated, under NASA contract NAS5-26555.  K.I.\"O  acknowledges the support of the Simons Foundation through a Simons Collaboration on the Origins of Life (SCOL) PI grant (No. 321183).  E.A.B. and K.I.\"O acknowledge support from NSF Grant \#1907653.  L.I.C. acknowledges support from the David and Lucille Packard Foundation and from the Virginia Space Grant Consortium.  J. H. acknowledges support from the National Science Foundation Graduate Research Fellowship under Grant No. DGE-1144152.  J.K.J acknowledges support from the H2020 European Research Council (ERC) (grant agreement No 646908) through ERC Consolidator Grant ``S4F''.  

\software{
{\fontfamily{qcr}\selectfont NumPy} \citep{VanDerWalt2011},
{\fontfamily{qcr}\selectfont Matplotlib} \citep{Hunter2007},
{\fontfamily{qcr}\selectfont Astropy} \citep{Astropy2013}, 
{\fontfamily{qcr}\selectfont SciPy} \citep{SciPy2020}, 
}

\FloatBarrier
\clearpage

\appendix 

\section{Line observation details}
\label{sec:app_lineobs} 

\begin{figure*}
\centering
	\includegraphics[width=0.5\linewidth]{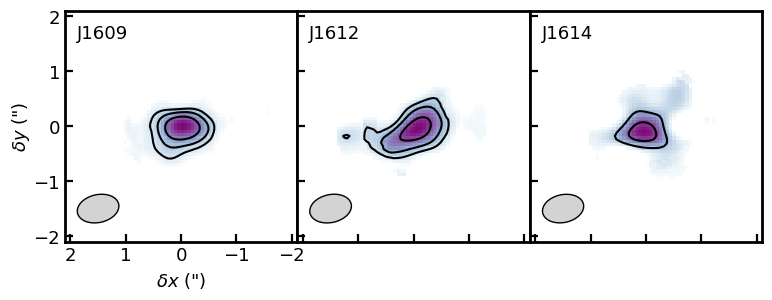}
	\caption{$^{13}$CO 2--1 moment zero maps.  Contours represent 3, 5, 8$\times$ the median rms across the map.  The rms values for each panel can be found in Table \ref{tab:classii_fluxes}.  Emission at a level below 1$\times$rms is not shown, and color scales are normalized to each individual image.  The restoring beams are shown in the lower left of each image.  }
\label{fig:mom0_13co}
\end{figure*}

Details on the line observations towards the Class II disks are listed in Table \ref{tab:classii_fluxes}.  Disk inclinations and position angles used for generating Keplerian masks and radial profiles are listed in Table \ref{tab:disk_geom}.  Moment zero maps for $^{13}$CO 2--1 observed towards J1609, J1612, and J1614 are shown in Figure \ref{fig:mom0_13co}.

\begin{deluxetable*}{lcccccc} 
	\tabletypesize{\footnotesize}
	\tablecaption{New Class II line observations \label{tab:classii_fluxes}}
	\tablecolumns{7} 
	\tablewidth{\textwidth} 
	\tablehead{
		\colhead{Transition}                          &
		\colhead{Beam dimensions}                       &
		\colhead{Chan. rms$^a$}                     & 
		\colhead{Vel. range$^b$}                    &
		\colhead{Mom. Zero rms$^c$}             &   
		\colhead{$\Omega^d$}                 &      
		\colhead{Mom. Zero flux$^d$} \\
		\colhead{}                                     & 
		\colhead{($\arcsec \times \arcsec$)}                               &
		\colhead{(mJy beam$^{-1}$)}                 & 
		\colhead{(km s$^{-1}$)}  &
		\colhead{(mJy beam$^{-1}$ km s$^{-1}$)}                 &
		\colhead{(square arcsec.)}         &
		\colhead{(mJy km s$^{-1}$)}               
		  }
\startdata
\hline 
 \multicolumn{7}{c}{CI Tau} \\ 
 \hline 
H$^{13}$CN 3-2 & 0.64 $\times$ 0.53 & 2.8 & 4.00--8.00 & 4.3 & 2.2 & $<$ 35\\
DCN 3-2 & 0.83 $\times$ 0.58 & 3.0 & 4.00--8.00 & 4.7 & 2.2 & 27 $\pm$ 10\\
HC$^{15}$N 3-2 & 0.64 $\times$ 0.53 & 2.8 & 4.00--8.00 & 4.0 & 2.2 & $<$ 31\\
\hline 
 \multicolumn{7}{c}{DM Tau} \\ 
 \hline 
H$^{13}$CN 3-2 & 0.60 $\times$ 0.52 & 2.8 & 3.50--8.50 & 4.2 & 3.4 & 58 $\pm$ 16\\
DCN 3-2 & 0.79 $\times$ 0.58 & 3.2 & 3.50--8.50 & 5.1 & 3.4 & $<$ 46\\
HC$^{15}$N 3-2 & 0.60 $\times$ 0.52 & 2.8 & 3.50--8.50 & 4.2 & 3.4 & $<$ 42\\
\hline 
 \multicolumn{7}{c}{DO Tau} \\ 
 \hline 
H$^{13}$CN 3-2 & 0.68 $\times$ 0.53 & 2.7 & 4.00--8.00 & 3.8 & 3.6 & $<$ 40\\
DCN 3-2 & 0.90 $\times$ 0.58 & 3.0 & 4.00--8.00 & 4.6 & 3.6 & $<$ 29\\
HC$^{15}$N 3-2 & 0.68 $\times$ 0.53 & 2.7 & 4.00--8.00 & 3.9 & 3.6 & $<$ 38\\
\hline 
 \multicolumn{7}{c}{HD 143006} \\ 
 \hline 
HCN 3-2 & 0.56 $\times$ 0.45 & 5.5 & 4.00--11.00 & 9.0 & 5.1 & 2149 $\pm$ 220\\
DCN 3-2 & 0.77 $\times$ 0.51 & 3.0 & 6.50--9.50 & 3.9 & 4.7 & 82 $\pm$ 15\\
\hline 
 \multicolumn{7}{c}{J1604} \\ 
 \hline 
HCN 3-2 & 0.41 $\times$ 0.33 & 7.7 & 1.50--7.00 & 10.8 & 11.8 & 8507 $\pm$ 850\\
DCN 3-2 & 0.51 $\times$ 0.36 & 4.8 & 4.00--5.50 & 4.2 & 4.7 & 295 $\pm$ 31\\
\hline 
 \multicolumn{7}{c}{J1609} \\ 
 \hline 
HCN 3-2 & 0.58 $\times$ 0.47 & 5.4 & -1.00--9.00 & 11.0 & 3.0 & 1048 $\pm$ 110\\
DCN 3-2 & 0.80 $\times$ 0.53 & 3.0 & 1.00--7.00 & 4.9 & 3.6 & 67 $\pm$ 18\\
$^{13}$CO 2-1 & 0.76 $\times$ 0.51 & 3.9 & -1.00--9.00 & 7.9 & 3.6 & 118 $\pm$ 26\\
\hline 
 \multicolumn{7}{c}{J1612} \\ 
 \hline 
HCN 3-2 & 0.58 $\times$ 0.47 & 5.4 & 2.00--7.50 & 8.5 & 1.7 & $<$ 79\\
DCN 3-2 & 0.80 $\times$ 0.53 & 3.0 & 2.00--7.50 & 5.3 & 1.7 & $<$ 36\\
$^{13}$CO 2-1 & 0.76 $\times$ 0.51 & 3.9 & 2.00--7.50 & 6.1 & 2.7 & 103 $\pm$ 20\\
\hline 
 \multicolumn{7}{c}{J1614} \\ 
 \hline 
HCN 3-2 & 0.58 $\times$ 0.46 & 5.3 & -1.00--8.50 & 11.4 & 2.4 & 130 $\pm$ 40\\
DCN 3-2 & 0.79 $\times$ 0.53 & 3.0 & -1.00--8.50 & 6.6 & 2.4 & $<$ 51\\
$^{13}$CO 2-1 & 0.75 $\times$ 0.51 & 4.0 & 2.00--9.50 & 7.2 & 3.7 & 75 $\pm$ 22\\
\hline 
 \multicolumn{7}{c}{LkCa 15} \\ 
 \hline 
H$^{13}$CN 3-2 & 0.64 $\times$ 0.52 & 2.7 & 3.00--9.00 & 3.4 & 6.5 & 202 $\pm$ 26\\
HC$^{15}$N 3-2 & 0.64 $\times$ 0.53 & 2.8 & 3.50--9.50 & 3.4 & 6.5 & 83 $\pm$ 21\\
\hline 
 \multicolumn{7}{c}{MWC 480} \\ 
 \hline 
H$^{13}$CN 3-2 & 0.77 $\times$ 0.50 & 3.9 & 0.00--8.50 & 6.8 & 4.2 & 176 $\pm$ 36\\
HC$^{15}$N 3-2 & 0.77 $\times$ 0.50 & 3.9 & 1.50--10.50 & 7.2 & 4.2 & 82 $\pm$ 25\\
\enddata
\tablenotetext{}{$^a$ For 0.5 km s$^{-1}$ channels.  $^b$ Velocity range included in the moment zero map.  $^c$ Median moment zero map rms. $^d$ Fluxes are extracted by summing the moment zero map, which has an angular size $\Omega$.  3$\sigma$ upper limits are reported for nondetections.  Uncertainties are derived by bootstrapping, added in quadrature with a 10\% calibration uncertainty.}
\end{deluxetable*}

\begin{deluxetable*}{lccc} 
	\tabletypesize{\footnotesize}
	\tablecaption{Class II disk geometries \label{tab:disk_geom}}
	\tablecolumns{4} 
	\tablewidth{\textwidth}     
	\tablehead{
	\colhead{Source}                          &
	\colhead{Inclination ($^{\circ}$)}                          &
	\colhead{Position angle ($^{\circ}$)}                          &
	\colhead{Reference}                          
}                
\startdata
AS 209               & 38 & 86 & 1 \\
CI Tau                & 50 & 11 & 2 \\
DM Tau              &34 & 155 & 3 \\
DO Tau              & 28 & 170 & 2 \\
HD 143006        & 17 & 170 & 4 \\
HD 163296        & 48 & 132 & 1 \\
IM Lup               & 50 & 144 & 1 \\
J1604-2130       & 6 & 80 & 5 \\
J1609-1908       & 50 & 95 & 6 \\
J1612-1859       & 53 & 104 & 6\\
J1614-1906       & 27 & 19 & 6 \\
LkCa 15            & 52 & 60 & 1 \\
MWC 480          & 37 & 148 & 1 \\
V4046 Sgr         & 33 & 76 & 1 \\ 
\enddata
\tablenotetext{}{References: [1] \citet{Huang2017}, [2] \citet{Long2019}, [3] \citet{Teague2015}, [4] \citet{Benisty2018}, 
	[5] \citet{Dong2017},  [6] \citet{Barenfeld2017}}.  
\end{deluxetable*}

\FloatBarrier 
\section{Column density estimates}
\label{sec:app_cd} 

C$_2$H, HCN, and C$^{18}$O column density estimates for the Class II sources are shown in Table \ref{tab:classii_cd}.  Continuum fluxes and H$_2$ column density estimates for all sources are listed in Table \ref{tab:cont_h2}.

\begin{deluxetable*}{lrcrr} 
	\tabletypesize{\footnotesize}
	\tablecaption{Class II line fluxes and column densities within the 4$\sigma$ emission radius \label{tab:classii_cd}}
	\tablecolumns{5} 
	\tablewidth{\textwidth} 
	\tablehead{
		\colhead{Source}                          &
		\colhead{Flux  $^a$}                       &
		\colhead{$\Omega$  $^a$}                     & 
		\colhead{$N_T$}                    &
		\colhead{$N_T$ ($\tau$-corrected)$^b$}                      \\
		\colhead{}                                     & 
		\colhead{(mJy km s$^{-1}$)}                               &
		\colhead{(square arcsec.)}                 & 
		\colhead{(cm$^{-2}$)}  &
		\colhead{(cm$^{-2}$)}                                                                  
		  }
\startdata
\hline 
 \multicolumn{5}{c}{C$^{18}$O} \\ 
 \hline 
AS 209 & 382 $\pm$ 47 & 8.0 & 8.5 $\pm$ 2.2 $\times$10$^{14}$ & \\
CI Tau & 607 $\pm$ 65 & 26.1 & 4.1 $\pm$ 1.0 $\times$10$^{14}$ & \\
DM Tau & 1119 $\pm$ 110 & 40.8 & 4.9 $\pm$ 1.2 $\times$10$^{14}$ & \\
DO Tau & 246 $\pm$ 29 & 5.7 & 7.6 $\pm$ 1.9 $\times$10$^{14}$ & \\
HD 143006 & 129 $\pm$ 17 & 3.1 & 7.4 $\pm$ 2.0 $\times$10$^{14}$ & \\
HD 163296 & 5925 $\pm$ 590 & 37.1 & 2.8 $\pm$ 0.7 $\times$10$^{15}$ & \\
IM Lup & 1233 $\pm$ 120 & 19.7 & 1.1 $\pm$ 0.3 $\times$10$^{15}$ & \\
J1604 & 1365 $\pm$ 140 & 10.6 & 2.3 $\pm$ 0.6 $\times$10$^{15}$ & \\
LkCa 15 & 414 $\pm$ 45 & 8.8 & 8.3 $\pm$ 2.1 $\times$10$^{14}$ & \\
MWC 480 & 1462 $\pm$ 150 & 15.5 & 1.7 $\pm$ 0.4 $\times$10$^{15}$ & \\
V4046 Sgr & 1179 $\pm$ 120 & 25.4 & 8.3 $\pm$ 2.1 $\times$10$^{14}$ & \\
\hline 
 \multicolumn{5}{c}{$^{13}$CO} \\ 
 \hline 
J1609 & 97 $\pm$ 13 & 1.0 & 1.7 $\pm$ 0.5 $\times$10$^{15}$ & \\
J1612 & 83 $\pm$ 11 & 1.3 & 1.1 $\pm$ 0.3 $\times$10$^{15}$ & \\
J1614 & 44 $\pm$ 10 & 0.7 & 1.1 $\pm$ 0.3 $\times$10$^{15}$ & \\
\hline 
 \multicolumn{5}{c}{HCN} \\ 
 \hline 
AS 209 & 3678 $\pm$ 370 & 8.4 & 6.8 $\pm$ 1.0 $\times$10$^{12}$ & 4.0 $\pm$ 1.3 $\times$10$^{13}$\\
HD 143006 & 2118 $\pm$ 210 & 3.9 & 8.5 $\pm$ 1.2 $\times$10$^{12}$ & 5.0 $\pm$ 1.7 $\times$10$^{13}$\\
HD 163296 & 9532 $\pm$ 950 & 51.6 & 2.9 $\pm$ 0.4 $\times$10$^{12}$ & 1.7 $\pm$ 0.6 $\times$10$^{13}$\\
IM Lup & 5683 $\pm$ 570 & 21.8 & 4.1 $\pm$ 0.6 $\times$10$^{12}$ & 2.4 $\pm$ 0.8 $\times$10$^{13}$\\
J1604 & 8284 $\pm$ 830 & 9.8 & 1.3 $\pm$ 0.2 $\times$10$^{13}$ & 7.6 $\pm$ 2.6 $\times$10$^{13}$\\
J1609 & 1025 $\pm$ 110 & 2.6 & 6.1 $\pm$ 0.9 $\times$10$^{12}$ & 3.6 $\pm$ 1.2 $\times$10$^{13}$\\
J1612 & $<$ 26 & 0.5 & $<$ 8.3 $\times$10$^{11}$ & $<$ 4.8 $\times$10$^{12}$\\
J1614 & 50 $\pm$ 12 & 0.4 & 2.2 $\pm$ 0.6 $\times$10$^{12}$ & 1.3 $\pm$ 0.5 $\times$10$^{13}$\\
V4046 Sgr & 9883 $\pm$ 990 & 85.7 & 1.8 $\pm$ 0.3 $\times$10$^{12}$ & 1.0 $\pm$ 0.3 $\times$10$^{13}$\\
\hline 
 \multicolumn{5}{c}{H$^{13}$CN} \\ 
 \hline 
CI Tau & $<$ 30 & 2.2 & $<$ 2.3 $\times$10$^{11}$ & \\
DM Tau & 43 $\pm$ 11 & 2.1 & 3.5 $\pm$ 1.0 $\times$10$^{11}$ & \\
DO Tau & $<$ 21 & 1.2 & $<$ 3.0 $\times$10$^{11}$ & \\
LkCa 15 & 176 $\pm$ 24 & 5.4 & 5.5 $\pm$ 0.9 $\times$10$^{11}$ & \\
MWC 480 & 150 $\pm$ 24 & 1.7 & 1.5 $\pm$ 0.3 $\times$10$^{12}$ & \\
\hline 
 \multicolumn{5}{c}{C$_2$H} \\ 
 \hline 
AS 209 & 2800 $\pm$ 300 & 10.2 & 1.2 $\pm$ 0.2 $\times$10$^{14}$ & 1.9 $\pm$ 0.3 $\times$10$^{14}$\\
CI Tau & 988 $\pm$ 100 & 15.8 & 2.7 $\pm$ 0.4 $\times$10$^{13}$ & 4.4 $\pm$ 0.8 $\times$10$^{13}$\\
DM Tau & 2012 $\pm$ 200 & 60.5 & 1.4 $\pm$ 0.2 $\times$10$^{13}$ & 2.4 $\pm$ 0.4 $\times$10$^{13}$\\
DO Tau & $<$ 20 & 1.2 & $<$ 7.5 $\times$10$^{12}$ & $<$ 1.2 $\times$10$^{13}$\\
HD 143006 & 382 $\pm$ 49 & 3.8 & 4.3 $\pm$ 0.7 $\times$10$^{13}$ & 7.0 $\pm$ 1.3 $\times$10$^{13}$\\
HD 163296 & 4511 $\pm$ 450 & 47.8 & 4.1 $\pm$ 0.6 $\times$10$^{13}$ & 6.7 $\pm$ 1.1 $\times$10$^{13}$\\
IM Lup & 1240 $\pm$ 140 & 15.0 & 3.6 $\pm$ 0.6 $\times$10$^{13}$ & 5.9 $\pm$ 1.0 $\times$10$^{13}$\\
J1604 & 2598 $\pm$ 260 & 9.8 & 1.2 $\pm$ 0.2 $\times$10$^{14}$ & 1.9 $\pm$ 0.3 $\times$10$^{14}$\\
J1609 & 584 $\pm$ 64 & 2.5 & 1.0 $\pm$ 0.2 $\times$10$^{14}$ & 1.7 $\pm$ 0.3 $\times$10$^{14}$\\
J1612 & $<$ 32 & 0.5 & $<$ 3.0 $\times$10$^{13}$ & $<$ 4.9 $\times$10$^{13}$\\
J1614 & $<$ 27 & 0.4 & $<$ 2.6 $\times$10$^{13}$ & $<$ 4.3 $\times$10$^{13}$\\
LkCa 15 & 2401 $\pm$ 240 & 29.5 & 3.5 $\pm$ 0.5 $\times$10$^{13}$ & 5.8 $\pm$ 1.0 $\times$10$^{13}$\\
MWC 480 & 1823 $\pm$ 180 & 5.0 & 1.6 $\pm$ 0.2 $\times$10$^{14}$ & 2.6 $\pm$ 0.4 $\times$10$^{14}$\\
V4046 Sgr & 2938 $\pm$ 300 & 80.2 & 1.6 $\pm$ 0.2 $\times$10$^{13}$ & 2.6 $\pm$ 0.4 $\times$10$^{13}$\\
\enddata
\tablenotetext{}{$^a$ Integrated within the maximum radius containing emission at a 4$\sigma$ level, as described in Section \ref{sec:classii_molcd}.  3$\sigma$ upper limits are reported for nondetections.  Uncertainties are derived by bootstrapping, added in quadrature with a 10\% calibration uncertainty.  $^b$ With an optical depth correction applied as described in Section \ref{sec:classii_molcd}.}
\end{deluxetable*}

\begin{deluxetable*}{lccccc} 
	\tabletypesize{\footnotesize}
	\tablecaption{Continuum fluxes and H$_2$ column density estimates \label{tab:cont_h2}}
	\tablecolumns{5} 
	\tablewidth{\textwidth} 
	\tablehead{
		\colhead{Source}                          &
		\colhead{Beam dim.}                    &
		\colhead{$\Omega$}                     & 
		\colhead{260 GHz continuum flux }                       &
		\colhead{$N_{T}$(H$_2$)}                    \\
		\colhead{}                                        &
		\colhead{($\arcsec \times \arcsec$)} &
		\colhead{(square arcsec.)}                 & 
		\colhead{(mJy)}                               &
		\colhead{(cm$^{-2}$)}                                                                  
		  }
\startdata
Ser-emb 1 & 0.50 $\times$ 0.42 & 0.4  & 127.3 & 8.5 $\times$10$^{24}$ \\
Ser-emb 7 & 0.50 $\times$ 0.42 & 0.2  & 1.3 & 1.6 $\times$10$^{23}$ \\
Ser-emb 8 & 0.50 $\times$ 0.42 & 0.4  & 53.3 & 3.3 $\times$10$^{24}$ \\
Ser-emb 15 & 0.50 $\times$ 0.42 & 0.4  & 20.8 & 1.3 $\times$10$^{24}$ \\
Ser-emb 17 & 0.50 $\times$ 0.42 & 0.4  & 78.8 & 5.5 $\times$10$^{24}$ \\
AS 209 & 0.53 $\times$ 0.50 & 3.9  & 277.3 & 1.9 $\times$10$^{24}$ \\
CI Tau & 0.52 $\times$ 0.43 & 2.7  & 173.9 & 1.7 $\times$10$^{24}$ \\
DM Tau & 0.49 $\times$ 0.43 & 7.1  & 131.6 & 5.0 $\times$10$^{23}$ \\
DO Tau & 0.56 $\times$ 0.43 & 1.4  & 151.2 & 2.9 $\times$10$^{24}$ \\
HD 143006 & 0.49 $\times$ 0.39 & 1.5  & 78.4 & 1.4 $\times$10$^{24}$ \\
HD 163296 & 0.46 $\times$ 0.40 & 10.2  & 877.1 & 2.3 $\times$10$^{24}$ \\
IM Lup & 0.50 $\times$ 0.48 & 8.8  & 453.5 & 1.4 $\times$10$^{24}$ \\
J1604 & 0.48 $\times$ 0.38 & 4.2  & 111.5 & 7.2 $\times$10$^{23}$ \\
J1609 & 0.48 $\times$ 0.38 & 0.6  & 26.1 & 1.2 $\times$10$^{24}$ \\
J1612 & 0.48 $\times$ 0.38 & 0.6  & 3.9 & 1.8 $\times$10$^{23}$ \\
J1614 & 0.49 $\times$ 0.38 & 0.5  & 21.5 & 1.1 $\times$10$^{24}$ \\
LkCa 15 & 0.52 $\times$ 0.43 & 2.3  & 168.4 & 1.9 $\times$10$^{24}$ \\
MWC 480 & 0.67 $\times$ 0.42 & 5.1  & 375.0 & 2.0 $\times$10$^{24}$ \\
V4046 Sgr & 0.77 $\times$ 0.56 & 3.2  & 218.8 & 1.9 $\times$10$^{24}$ \\
\enddata
\tablenotetext{}{Class 0/I continuum fluxes are measured within the DCN mask used for line extractions (Section \ref{sec:class0_structures}).  Class II continuum fluxes are measured within an ellipse fit to the 3--5$\sigma$ continuum contour, depending on the SNR.  See Section \ref{sec:caveats} for a discussion of uncertainty sources in deriving H$_2$ column densities.}
\end{deluxetable*}

\FloatBarrier
\clearpage
\bibliography{references}

\begin{thebibliography}{}
\expandafter\ifx\csname natexlab\endcsname\relax\def\natexlab#1{#1}\fi
\providecommand{\url}[1]{\href{#1}{#1}}
\providecommand{\dodoi}[1]{doi:~\href{http://doi.org/#1}{\nolinkurl{#1}}}
\providecommand{\doeprint}[1]{\href{http://ascl.net/#1}{\nolinkurl{http://ascl.net/#1}}}
\providecommand{\doarXiv}[1]{\href{https://arxiv.org/abs/#1}{\nolinkurl{https://arxiv.org/abs/#1}}}

\bibitem[{{Ahrens} {et~al.}(2002){Ahrens}, {Lewen}, {Takano}, {Winnewisser},
  {Urban}, {Negirev}, \& {Koroliev}}]{Ahrens2002}
{Ahrens}, V., {Lewen}, F., {Takano}, S., {et~al.} 2002, Zeitschrift
  Naturforschung Teil A, 57, 669, \dodoi{10.1515/zna-2002-0806}

\bibitem[{{Aikawa} \& {Herbst}(1999)}]{Aikawa1999}
{Aikawa}, Y., \& {Herbst}, E. 1999, \aap, 351, 233

\bibitem[{{ALMA Partnership} {et~al.}(2015){ALMA Partnership}, {Brogan},
  {P{\'e}rez}, {Hunter}, {Dent}, {Hales}, {Corder}, {Fomalont}, {Vlahakis},
  {Asaki}, {Barkats}, {Hirota}, {Hodge}, {Impellizzeri}, {Kneissl}, {Liuzzo},
  {Lucas}, {Marcelino}, {Matsushita}, {Nakanishi}, {Phillips}, {Richards},
  {Toledo}, {Aladro}, {Broguiere}, {Cortes}, {Cortes}, {Espada}, {Galarza},
  {Garcia-Appadoo}, {Guzman-Ramirez}, {Humphreys}, {Jung}, {Kameno}, {Laing},
  {Leon}, {Marconi}, {Mignano}, {Nikolic}, {Nyman}, {Radiszcz}, {Remijan},
  {Rod{\'o}n}, {Sawada}, {Takahashi}, {Tilanus}, {Vila Vilaro}, {Watson},
  {Wiklind}, {Akiyama}, {Chapillon}, {de Gregorio-Monsalvo}, {Di Francesco},
  {Gueth}, {Kawamura}, {Lee}, {Nguyen Luong}, {Mangum}, {Pietu}, {Sanhueza},
  {Saigo}, {Takakuwa}, {Ubach}, {van Kempen}, {Wootten}, {Castro-Carrizo},
  {Francke}, {Gallardo}, {Garcia}, {Gonzalez}, {Hill}, {Kaminski}, {Kurono},
  {Liu}, {Lopez}, {Morales}, {Plarre}, {Schieven}, {Testi}, {Videla},
  {Villard}, {Andreani}, {Hibbard}, \& {Tatematsu}}]{ALMA2015}
{ALMA Partnership}, {Brogan}, C.~L., {P{\'e}rez}, L.~M., {et~al.} 2015, \apjl,
  808, L3, \dodoi{10.1088/2041-8205/808/1/L3}

\bibitem[{{Anderl} {et~al.}(2016){Anderl}, {Maret}, {Cabrit}, {Belloche},
  {Maury}, {Andr{\'e}}, {Codella}, {Bacmann}, {Bontemps}, {Podio}, {Gueth}, \&
  {Bergin}}]{Anderl2016}
{Anderl}, S., {Maret}, S., {Cabrit}, S., {et~al.} 2016, \aap, 591, A3,
  \dodoi{10.1051/0004-6361/201527831}

\bibitem[{{Anderson} {et~al.}(2017){Anderson}, {Bergin}, {Blake}, {Ciesla},
  {Visser}, \& {Lee}}]{Anderson2017}
{Anderson}, D.~E., {Bergin}, E.~A., {Blake}, G.~A., {et~al.} 2017, \apj, 845,
  13, \dodoi{10.3847/1538-4357/aa7da1}

\bibitem[{{Andrews} \& {Williams}(2005)}]{Andrews2005}
{Andrews}, S.~M., \& {Williams}, J.~P. 2005, \apj, 631, 1134,
  \dodoi{10.1086/432712}

\bibitem[{{Andrews} {et~al.}(2018){Andrews}, {Huang}, {P{\'e}rez}, {Isella},
  {Dullemond}, {Kurtovic}, {Guzm{\'a}n}, {Carpenter}, {Wilner}, {Zhang}, {Zhu},
  {Birnstiel}, {Bai}, {Benisty}, {Hughes}, {{\"O}berg}, \&
  {Ricci}}]{Andrews2018}
{Andrews}, S.~M., {Huang}, J., {P{\'e}rez}, L.~M., {et~al.} 2018, \apjl, 869,
  L41, \dodoi{10.3847/2041-8213/aaf741}

\bibitem[{{Artur de la Villarmois} {et~al.}(2019){Artur de la Villarmois},
  {Kristensen}, \& {J{\o}rgensen}}]{Artur2019}
{Artur de la Villarmois}, E., {Kristensen}, L.~E., \& {J{\o}rgensen}, J.~K.
  2019, \aap, 627, A37, \dodoi{10.1051/0004-6361/201935575}

\bibitem[{{Astropy Collaboration} {et~al.}(2013){Astropy Collaboration},
  {Robitaille}, {Tollerud}, {Greenfield}, {Droettboom}, {Bray}, {Aldcroft},
  {Davis}, {Ginsburg}, {Price-Whelan}, {Kerzendorf}, {Conley}, {Crighton},
  {Barbary}, {Muna}, {Ferguson}, {Grollier}, {Parikh}, {Nair}, {Unther},
  {Deil}, {Woillez}, {Conseil}, {Kramer}, {Turner}, {Singer}, {Fox}, {Weaver},
  {Zabalza}, {Edwards}, {Azalee Bostroem}, {Burke}, {Casey}, {Crawford},
  {Dencheva}, {Ely}, {Jenness}, {Labrie}, {Lim}, {Pierfederici}, {Pontzen},
  {Ptak}, {Refsdal}, {Servillat}, \& {Streicher}}]{Astropy2013}
{Astropy Collaboration}, {Robitaille}, T.~P., {Tollerud}, E.~J., {et~al.} 2013,
  \aap, 558, A33, \dodoi{10.1051/0004-6361/201322068}

\bibitem[{{Barenfeld} {et~al.}(2017){Barenfeld}, {Carpenter}, {Sargent},
  {Isella}, \& {Ricci}}]{Barenfeld2017}
{Barenfeld}, S.~A., {Carpenter}, J.~M., {Sargent}, A.~I., {Isella}, A., \&
  {Ricci}, L. 2017, \apj, 851, 85, \dodoi{10.3847/1538-4357/aa989d}

\bibitem[{{Beckwith} {et~al.}(1990){Beckwith}, {Sargent}, {Chini}, \&
  {Guesten}}]{Beckwith1990}
{Beckwith}, S. V.~W., {Sargent}, A.~I., {Chini}, R.~S., \& {Guesten}, R. 1990,
  \aj, 99, 924, \dodoi{10.1086/115385}

\bibitem[{{Benisty} {et~al.}(2018){Benisty}, {Juh{\'a}sz}, {Facchini},
  {Pinilla}, {de Boer}, {P{\'e}rez}, {Keppler}, {Muro-Arena}, {Villenave},
  {Andrews}, {Dominik}, {Dullemond}, {Gallenne}, {Garufi}, {Ginski}, \&
  {Isella}}]{Benisty2018}
{Benisty}, M., {Juh{\'a}sz}, A., {Facchini}, S., {et~al.} 2018, \aap, 619,
  A171, \dodoi{10.1051/0004-6361/201833913}

\bibitem[{{Bergin} {et~al.}(2016){Bergin}, {Du}, {Cleeves}, {Blake}, {Schwarz},
  {Visser}, \& {Zhang}}]{Bergin2016}
{Bergin}, E.~A., {Du}, F., {Cleeves}, L.~I., {et~al.} 2016, \apj, 831, 101,
  \dodoi{10.3847/0004-637X/831/1/101}

\bibitem[{{Bergin} {et~al.}(2010){Bergin}, {Hogerheijde}, {Brinch}, {Fogel},
  {Y{\i}ld{\i}z}, {Kristensen}, {van Dishoeck}, {Bell}, {Blake}, {Cernicharo},
  {Dominik}, {Lis}, {Melnick}, {Neufeld}, {Pani{\'c}}, {Pearson}, {Bachiller},
  {Baudry}, {Benedettini}, {Benz}, {Bjerkeli}, {Bontemps}, {Braine},
  {Bruderer}, {Caselli}, {Codella}, {Daniel}, {di Giorgio}, {Doty}, {Encrenaz},
  {Fich}, {Fuente}, {Giannini}, {Goicoechea}, {de Graauw}, {Helmich},
  {Herczeg}, {Herpin}, {Jacq}, {Johnstone}, {J{\o}rgensen}, {Larsson},
  {Liseau}, {Marseille}, {McCoey}, {Nisini}, {Olberg}, {Parise}, {Plume},
  {Risacher}, {Santiago-Garc{\'\i}a}, {Saraceno}, {Shipman}, {Tafalla}, {van
  Kempen}, {Visser}, {Wampfler}, {Wyrowski}, {van der Tak}, {Jellema},
  {Tielens}, {Hartogh}, {St{\"u}tzki}, \& {Szczerba}}]{Bergin2010}
{Bergin}, E.~A., {Hogerheijde}, M.~R., {Brinch}, C., {et~al.} 2010, \aap, 521,
  L33, \dodoi{10.1051/0004-6361/201015104}

\bibitem[{{Bergin} {et~al.}(2013){Bergin}, {Cleeves}, {Gorti}, {Zhang},
  {Blake}, {Green}, {Andrews}, {Evans}, {Henning}, {{\"O}berg}, {Pontoppidan},
  {Qi}, {Salyk}, \& {van Dishoeck}}]{Bergin2013}
{Bergin}, E.~A., {Cleeves}, L.~I., {Gorti}, U., {et~al.} 2013, \nat, 493, 644,
  \dodoi{10.1038/nature11805}

\bibitem[{{Bergner} {et~al.}(2018){Bergner}, {Guzm{\'a}n}, {{\"O}berg},
  {Loomis}, \& {Pegues}}]{Bergner2018}
{Bergner}, J.~B., {Guzm{\'a}n}, V.~G., {{\"O}berg}, K.~I., {Loomis}, R.~A., \&
  {Pegues}, J. 2018, \apj, 857, 69, \dodoi{10.3847/1538-4357/aab664}

\bibitem[{{Bergner} {et~al.}(2019{\natexlab{a}}){Bergner},
  {Mart{\'\i}n-Dom{\'e}nech}, {{\"O}berg}, {J{\o}rgensen}, {Artur de la
  Villarmois}, \& {Brinch}}]{Bergner2019c}
{Bergner}, J.~B., {Mart{\'\i}n-Dom{\'e}nech}, R., {{\"O}berg}, K.~I., {et~al.}
  2019{\natexlab{a}}, ACS Earth and Space Chemistry, 3, 1564,
  \dodoi{10.1021/acsearthspacechem.9b00059}

\bibitem[{{Bergner} {et~al.}(2019{\natexlab{b}}){Bergner}, {{\"O}berg},
  {Bergin}, {Loomis}, {Pegues}, \& {Qi}}]{Bergner2019b}
{Bergner}, J.~B., {{\"O}berg}, K.~I., {Bergin}, E.~A., {et~al.}
  2019{\natexlab{b}}, \apj, 876, 25, \dodoi{10.3847/1538-4357/ab141e}

\bibitem[{{Bockel{\'e}e-Morvan} {et~al.}(2008){Bockel{\'e}e-Morvan}, {Biver},
  {Jehin}, {Cochran}, {Wiesemeyer}, {Manfroid}, {Hutsem{\'e}kers}, {Arpigny},
  {Boissier}, {Cochran}, {Colom}, {Crovisier}, {Milutinovic}, {Moreno},
  {Prochaska}, {Ramirez}, {Schulz}, \& {Zucconi}}]{Bockelee-Morvan2008}
{Bockel{\'e}e-Morvan}, D., {Biver}, N., {Jehin}, E., {et~al.} 2008, \apjl, 679,
  L49, \dodoi{10.1086/588781}

\bibitem[{{Booth} {et~al.}(2019){Booth}, {Walsh}, {Ilee}, {Notsu}, {Qi},
  {Nomura}, \& {Akiyama}}]{Booth2019}
{Booth}, A.~S., {Walsh}, C., {Ilee}, J.~D., {et~al.} 2019, \apjl, 882, L31,
  \dodoi{10.3847/2041-8213/ab3645}

\bibitem[{{Br{\"u}nken} {et~al.}(2004){Br{\"u}nken}, {Fuchs}, {Lewen}, {Urban},
  {Giesen}, \& {Winnewisser}}]{Brunken2004}
{Br{\"u}nken}, S., {Fuchs}, U., {Lewen}, F., {et~al.} 2004, Journal of
  Molecular Spectroscopy, 225, 152, \dodoi{10.1016/j.jms.2004.02.021}

\bibitem[{{Cazzoli} \& {Puzzarini}(2005)}]{Cazzoli2005b}
{Cazzoli}, G., \& {Puzzarini}, C. 2005, Journal of Molecular Spectroscopy, 233,
  280, \dodoi{10.1016/j.jms.2005.07.009}

\bibitem[{{Cazzoli} {et~al.}(2005){Cazzoli}, {Puzzarini}, \&
  {Gauss}}]{Cazzoli2005a}
{Cazzoli}, G., {Puzzarini}, C., \& {Gauss}, J. 2005, \apjs, 159, 181,
  \dodoi{10.1086/430209}

\bibitem[{{Chapillon} {et~al.}(2008){Chapillon}, {Guilloteau}, {Dutrey}, \&
  {Pi{\'e}tu}}]{Chapillon2008}
{Chapillon}, E., {Guilloteau}, S., {Dutrey}, A., \& {Pi{\'e}tu}, V. 2008, \aap,
  488, 565, \dodoi{10.1051/0004-6361:200809523}

\bibitem[{{Cleeves} {et~al.}(2016){Cleeves}, {{\"O}berg}, {Wilner}, {Huang},
  {Loomis}, {Andrews}, \& {Czekala}}]{Cleeves2016}
{Cleeves}, L.~I., {{\"O}berg}, K.~I., {Wilner}, D.~J., {et~al.} 2016, \apj,
  832, 110, \dodoi{10.3847/0004-637X/832/2/110}

\bibitem[{{Cleeves} {et~al.}(2018){Cleeves}, {{\"O}berg}, {Wilner}, {Huang},
  {Loomis}, {Andrews}, \& {Guzman}}]{Cleeves2018}
---. 2018, \apj, 865, 155, \dodoi{10.3847/1538-4357/aade96}

\bibitem[{{Codella} {et~al.}(2019){Codella}, {Ceccarelli}, {Lee}, {Bianchi},
  {Balucani}, {Podio}, {Cabrit}, {Gueth}, {Gusdorf}, {Lefloch}, \&
  {Tabone}}]{Codella2019}
{Codella}, C., {Ceccarelli}, C., {Lee}, C.-F., {et~al.} 2019, ACS Earth and
  Space Chemistry, 3, 2110, \dodoi{10.1021/acsearthspacechem.9b00136}

\bibitem[{{Cridland} {et~al.}(2016){Cridland}, {Pudritz}, \&
  {Alessi}}]{Cridland2016}
{Cridland}, A.~J., {Pudritz}, R.~E., \& {Alessi}, M. 2016, \mnras, 461, 3274,
  \dodoi{10.1093/mnras/stw1511}

\bibitem[{{Dong} {et~al.}(2017){Dong}, {van der Marel}, {Hashimoto}, {Chiang},
  {Akiyama}, {Liu}, {Muto}, {Knapp}, {Tsukagoshi}, {Brown}, {Bruderer},
  {Koyamatsu}, {Kudo}, {Ohashi}, {Rich}, {Satoshi}, {Takami}, {Wisniewski},
  {Yang}, {Zhu}, \& {Tamura}}]{Dong2017}
{Dong}, R., {van der Marel}, N., {Hashimoto}, J., {et~al.} 2017, \apj, 836,
  201, \dodoi{10.3847/1538-4357/aa5abf}

\bibitem[{{Drozdovskaya} {et~al.}(2014){Drozdovskaya}, {Walsh}, {Visser},
  {Harsono}, \& {van Dishoeck}}]{Drozdovskaya2014}
{Drozdovskaya}, M.~N., {Walsh}, C., {Visser}, R., {Harsono}, D., \& {van
  Dishoeck}, E.~F. 2014, \mnras, 445, 913, \dodoi{10.1093/mnras/stu1789}

\bibitem[{{Du} {et~al.}(2015){Du}, {Bergin}, \& {Hogerheijde}}]{Du2015}
{Du}, F., {Bergin}, E.~A., \& {Hogerheijde}, M.~R. 2015, \apjl, 807, L32,
  \dodoi{10.1088/2041-8205/807/2/L32}

\bibitem[{{Du} {et~al.}(2017){Du}, {Bergin}, {Hogerheijde}, {van Dishoeck},
  {Blake}, {Bruderer}, {Cleeves}, {Dominik}, {Fedele}, {Lis}, {Melnick},
  {Neufeld}, {Pearson}, \& {Y{\i}ld{\i}z}}]{Du2017}
{Du}, F., {Bergin}, E.~A., {Hogerheijde}, M., {et~al.} 2017, \apj, 842, 98,
  \dodoi{10.3847/1538-4357/aa70ee}

\bibitem[{{Dunham} {et~al.}(2014{\natexlab{a}}){Dunham}, {Vorobyov}, \&
  {Arce}}]{Dunham2014b}
{Dunham}, M.~M., {Vorobyov}, E.~I., \& {Arce}, H.~G. 2014{\natexlab{a}},
  \mnras, 444, 887, \dodoi{10.1093/mnras/stu1511}

\bibitem[{{Dunham} {et~al.}(2014{\natexlab{b}}){Dunham}, {Stutz}, {Allen},
  {Evans}, {Fischer}, {Megeath}, {Myers}, {Offner}, {Poteet}, {Tobin}, \&
  {Vorobyov}}]{Dunham2014}
{Dunham}, M.~M., {Stutz}, A.~M., {Allen}, L.~E., {et~al.} 2014{\natexlab{b}},
  in Protostars and Planets VI, ed. H.~{Beuther}, R.~S. {Klessen}, C.~P.
  {Dullemond}, \& T.~{Henning}, 195

\bibitem[{{Dutrey} {et~al.}(2003){Dutrey}, {Guilloteau}, \&
  {Simon}}]{Dutrey2003}
{Dutrey}, A., {Guilloteau}, S., \& {Simon}, M. 2003, \aap, 402, 1003,
  \dodoi{10.1051/0004-6361:20030317}

\bibitem[{{Ehrenfreund} \& {Charnley}(2000)}]{Ehrenfreund2000}
{Ehrenfreund}, P., \& {Charnley}, S.~B. 2000, \araa, 38, 427,
  \dodoi{10.1146/annurev.astro.38.1.427}

\bibitem[{{Eistrup} {et~al.}(2018){Eistrup}, {Walsh}, \& {van
  Dishoeck}}]{Eistrup2018}
{Eistrup}, C., {Walsh}, C., \& {van Dishoeck}, E.~F. 2018, \aap, 613, A14,
  \dodoi{10.1051/0004-6361/201731302}

\bibitem[{{Enoch} {et~al.}(2009){Enoch}, {Evans}, {Sargent}, \&
  {Glenn}}]{Enoch2009}
{Enoch}, M.~L., {Evans}, II, N.~J., {Sargent}, A.~I., \& {Glenn}, J. 2009,
  \apj, 692, 973, \dodoi{10.1088/0004-637X/692/2/973}

\bibitem[{{Enoch} {et~al.}(2011){Enoch}, {Corder}, {Duch{\^e}ne}, {Bock},
  {Bolatto}, {Culverhouse}, {Kwon}, {Lamb}, {Leitch}, {Marrone}, {Muchovej},
  {P{\'e}rez}, {Scott}, {Teuben}, {Wright}, \& {Zauderer}}]{Enoch2011}
{Enoch}, M.~L., {Corder}, S., {Duch{\^e}ne}, G., {et~al.} 2011, \apjs, 195, 21,
  \dodoi{10.1088/0067-0049/195/2/21}

\bibitem[{{Evans} {et~al.}(2009){Evans}, {Dunham}, {J{\o}rgensen}, {Enoch},
  {Mer{\'\i}n}, {van Dishoeck}, {Alcal{\'a}}, {Myers}, {Stapelfeldt}, {Huard},
  {Allen}, {Harvey}, {van Kempen}, {Blake}, {Koerner}, {Mundy}, {Padgett}, \&
  {Sargent}}]{Evans2009}
{Evans}, Neal~J., I., {Dunham}, M.~M., {J{\o}rgensen}, J.~K., {et~al.} 2009,
  \apjs, 181, 321, \dodoi{10.1088/0067-0049/181/2/321}

\bibitem[{{Favre} {et~al.}(2015){Favre}, {Bergin}, {Cleeves}, {Hersant}, {Qi},
  \& {Aikawa}}]{Favre2015}
{Favre}, C., {Bergin}, E.~A., {Cleeves}, L.~I., {et~al.} 2015, \apjl, 802, L23,
  \dodoi{10.1088/2041-8205/802/2/L23}

\bibitem[{{Favre} {et~al.}(2013){Favre}, {Cleeves}, {Bergin}, {Qi}, \&
  {Blake}}]{Favre2013}
{Favre}, C., {Cleeves}, L.~I., {Bergin}, E.~A., {Qi}, C., \& {Blake}, G.~A.
  2013, \apjl, 776, L38, \dodoi{10.1088/2041-8205/776/2/L38}

\bibitem[{{Fayolle} {et~al.}(2016){Fayolle}, {Balfe}, {Loomis}, {Bergner},
  {Graninger}, {Rajappan}, \& {{\"O}berg}}]{Fayolle2016}
{Fayolle}, E.~C., {Balfe}, J., {Loomis}, R., {et~al.} 2016, \apjl, 816, L28,
  \dodoi{10.3847/2041-8205/816/2/L28}

\bibitem[{{Feng} {et~al.}(2019){Feng}, {Caselli}, {Wang}, {Lin}, {Beuther}, \&
  {Sipil{\"a}}}]{Feng2019}
{Feng}, S., {Caselli}, P., {Wang}, K., {et~al.} 2019, \apj, 883, 202,
  \dodoi{10.3847/1538-4357/ab3a42}

\bibitem[{{Foreman-Mackey} {et~al.}(2013){Foreman-Mackey}, {Hogg}, {Lang}, \&
  {Goodman}}]{Foreman2013}
{Foreman-Mackey}, D., {Hogg}, D.~W., {Lang}, D., \& {Goodman}, J. 2013, \pasp,
  125, 306, \dodoi{10.1086/670067}

\bibitem[{{Fuchs} {et~al.}(2004){Fuchs}, {Bruenken}, {Fuchs}, {Thorwirth},
  {Ahrens}, {Lewen}, {Urban}, {Giesen}, \& {Winnewisser}}]{Fuchs2004}
{Fuchs}, U., {Bruenken}, S., {Fuchs}, G.~W., {et~al.} 2004, Zeitschrift
  Naturforschung Teil A, 59, 861, \dodoi{10.1515/zna-2004-1123}

\bibitem[{{Gaia Collaboration} {et~al.}(2018){Gaia Collaboration}, {Brown},
  {Vallenari}, {Prusti}, {de Bruijne}, {Babusiaux}, {Bailer-Jones}, {Biermann},
  {Evans}, {Eyer}, {Jansen}, {Jordi}, {Klioner}, {Lammers}, {Lindegren},
  {Luri}, \& {Mignard}}]{Gaia2018}
{Gaia Collaboration}, {Brown}, A.~G.~A., {Vallenari}, A., {et~al.} 2018, \aap,
  616, A1, \dodoi{10.1051/0004-6361/201833051}

\bibitem[{{Geiss} \& {Gloeckler}(2003)}]{Geiss2003}
{Geiss}, J., \& {Gloeckler}, G. 2003, \ssr, 106, 3,
  \dodoi{10.1023/A:1024651232758}

\bibitem[{{Goldsmith} \& {Langer}(1999)}]{Goldsmith1999}
{Goldsmith}, P.~F., \& {Langer}, W.~D. 1999, \apj, 517, 209,
  \dodoi{10.1086/307195}

\bibitem[{{Guzm{\'a}n} {et~al.}(2017){Guzm{\'a}n}, {{\"O}berg}, {Huang},
  {Loomis}, \& {Qi}}]{Guzman2017}
{Guzm{\'a}n}, V.~V., {{\"O}berg}, K.~I., {Huang}, J., {Loomis}, R., \& {Qi}, C.
  2017, \apj, 836, 30, \dodoi{10.3847/1538-4357/836/1/30}

\bibitem[{{Harsono} {et~al.}(2018){Harsono}, {Bjerkeli}, {van der Wiel},
  {Ramsey}, {Maud}, {Kristensen}, \& {J{\o}rgensen}}]{Harsono2018}
{Harsono}, D., {Bjerkeli}, P., {van der Wiel}, M. H.~D., {et~al.} 2018, Nature
  Astronomy, 2, 646, \dodoi{10.1038/s41550-018-0497-x}

\bibitem[{{Harsono} {et~al.}(2020){Harsono}, {Persson}, {Ramos}, {Murillo},
  {Maud}, {Hogerheijde}, {Bosman}, {Kristensen}, {Jorgensen}, {Bergin},
  {Visser}, {Mottram}, \& {van Dishoeck}}]{Harsono2020}
{Harsono}, D., {Persson}, M., {Ramos}, A., {et~al.} 2020, arXiv e-prints,
  arXiv:2002.11897.
\newblock \doarXiv{2002.11897}

\bibitem[{{Heays} {et~al.}(2014){Heays}, {Visser}, {Gredel}, {Ubachs}, {Lewis},
  {Gibson}, \& {van Dishoeck}}]{Heays2014}
{Heays}, A.~N., {Visser}, R., {Gredel}, R., {et~al.} 2014, \aap, 562, A61,
  \dodoi{10.1051/0004-6361/201322832}

\bibitem[{{Hildebrand}(1983)}]{Hildebrand1983}
{Hildebrand}, R.~H. 1983, \qjras, 24, 267

\bibitem[{{Hily-Blant} {et~al.}(2013){Hily-Blant}, {Bonal}, {Faure}, \&
  {Quirico}}]{Hily-Blant2013}
{Hily-Blant}, P., {Bonal}, L., {Faure}, A., \& {Quirico}, E. 2013, \icarus,
  223, 582, \dodoi{10.1016/j.icarus.2012.12.015}

\bibitem[{{Hily-Blant} {et~al.}(2019){Hily-Blant}, {Magalhaes de Souza},
  {Kastner}, \& {Forveille}}]{Hily-Blant2019}
{Hily-Blant}, P., {Magalhaes de Souza}, V., {Kastner}, J., \& {Forveille}, T.
  2019, \aap, 632, L12, \dodoi{10.1051/0004-6361/201936750}

\bibitem[{{Hogerheijde} {et~al.}(2011){Hogerheijde}, {Bergin}, {Brinch},
  {Cleeves}, {Fogel}, {Blake}, {Dominik}, {Lis}, {Melnick}, {Neufeld},
  {Pani{\'c}}, {Pearson}, {Kristensen}, {Y{\i}ld{\i}z}, \& {van
  Dishoeck}}]{Hogerheijde2011}
{Hogerheijde}, M.~R., {Bergin}, E.~A., {Brinch}, C., {et~al.} 2011, Science,
  334, 338, \dodoi{10.1126/science.1208931}

\bibitem[{{Huang} {et~al.}(2017){Huang}, {{\"O}berg}, {Qi}, {Aikawa},
  {Andrews}, {Furuya}, {Guzm{\'a}n}, {Loomis}, {van Dishoeck}, \&
  {Wilner}}]{Huang2017}
{Huang}, J., {{\"O}berg}, K.~I., {Qi}, C., {et~al.} 2017, \apj, 835, 231,
  \dodoi{10.3847/1538-4357/835/2/231}

\bibitem[{{Hunter}(2007)}]{Hunter2007}
{Hunter}, J.~D. 2007, Computing in Science and Engineering, 9, 90,
  \dodoi{10.1109/MCSE.2007.55}

\bibitem[{{Ikeda} {et~al.}(2002){Ikeda}, {Hirota}, \& {Yamamoto}}]{Ikeda2002}
{Ikeda}, M., {Hirota}, T., \& {Yamamoto}, S. 2002, \apj, 575, 250,
  \dodoi{10.1086/341287}

\bibitem[{{J{\o}rgensen} {et~al.}(2002){J{\o}rgensen}, {Sch{\"o}ier}, \& {van
  Dishoeck}}]{Jorgensen2002}
{J{\o}rgensen}, J.~K., {Sch{\"o}ier}, F.~L., \& {van Dishoeck}, E.~F. 2002,
  \aap, 389, 908, \dodoi{10.1051/0004-6361:20020681}

\bibitem[{{J{\o}rgensen} {et~al.}(2004){J{\o}rgensen}, {Sch{\"o}ier}, \& {van
  Dishoeck}}]{Jorgensen2004}
---. 2004, \aap, 416, 603, \dodoi{10.1051/0004-6361:20034440}

\bibitem[{{J{\o}rgensen} {et~al.}(2005){J{\o}rgensen}, {Sch{\"o}ier}, \& {van
  Dishoeck}}]{Jorgensen2005}
---. 2005, \aap, 435, 177, \dodoi{10.1051/0004-6361:20042092}

\bibitem[{{J{\o}rgensen} {et~al.}(2009){J{\o}rgensen}, {van Dishoeck},
  {Visser}, {Bourke}, {Wilner}, {Lommen}, {Hogerheijde}, \&
  {Myers}}]{Jorgensen2009}
{J{\o}rgensen}, J.~K., {van Dishoeck}, E.~F., {Visser}, R., {et~al.} 2009,
  \aap, 507, 861, \dodoi{10.1051/0004-6361/200912325}

\bibitem[{{J{\o}rgensen} {et~al.}(2015){J{\o}rgensen}, {Visser}, {Williams}, \&
  {Bergin}}]{Jorgensen2015}
{J{\o}rgensen}, J.~K., {Visser}, R., {Williams}, J.~P., \& {Bergin}, E.~A.
  2015, \aap, 579, A23, \dodoi{10.1051/0004-6361/201425317}

\bibitem[{{Kerridge} {et~al.}(1987){Kerridge}, {Chang}, \&
  {Shipp}}]{Kerridge1987}
{Kerridge}, J.~F., {Chang}, S., \& {Shipp}, R. 1987, \gca, 51, 2527,
  \dodoi{10.1016/0016-7037(87)90303-6}

\bibitem[{{Klapper} {et~al.}(2001){Klapper}, {Lewen}, {Gendriesch}, {Belov}, \&
  {Winnewisser}}]{Klapper2001}
{Klapper}, G., {Lewen}, F., {Gendriesch}, R., {Belov}, S.~P., \& {Winnewisser},
  G. 2001, Zeitschrift Naturforschung Teil A, 56, 329,
  \dodoi{10.1515/zna-2001-0317}

\bibitem[{{Krijt} {et~al.}(2016){Krijt}, {Ciesla}, \& {Bergin}}]{Krijt2016}
{Krijt}, S., {Ciesla}, F.~J., \& {Bergin}, E.~A. 2016, \apj, 833, 285,
  \dodoi{10.3847/1538-4357/833/2/285}

\bibitem[{{Krijt} {et~al.}(2018){Krijt}, {Schwarz}, {Bergin}, \&
  {Ciesla}}]{Krijt2018}
{Krijt}, S., {Schwarz}, K.~R., {Bergin}, E.~A., \& {Ciesla}, F.~J. 2018, \apj,
  864, 78, \dodoi{10.3847/1538-4357/aad69b}

\bibitem[{{Lee} {et~al.}(2017){Lee}, {Li}, {Ho}, {Hirano}, {Zhang}, \&
  {Shang}}]{Lee2017}
{Lee}, C.-F., {Li}, Z.-Y., {Ho}, P. T.~P., {et~al.} 2017, \apj, 843, 27,
  \dodoi{10.3847/1538-4357/aa7757}

\bibitem[{{Linsky} {et~al.}(2006){Linsky}, {Draine}, {Moos}, {Jenkins}, {Wood},
  {Oliveira}, {Blair}, {Friedman}, {Gry}, {Knauth}, {Kruk}, {Lacour}, {Lehner},
  {Redfield}, {Shull}, {Sonneborn}, \& {Williger}}]{Linsky2006}
{Linsky}, J.~L., {Draine}, B.~T., {Moos}, H.~W., {et~al.} 2006, \apj, 647,
  1106, \dodoi{10.1086/505556}

\bibitem[{{Long} {et~al.}(2019){Long}, {Herczeg}, {Harsono}, {Pinilla},
  {Tazzari}, {Manara}, {Pascucci}, {Cabrit}, {Nisini}, {Johnstone}, {Edwards},
  {Salyk}, {Menard}, {Lodato}, {Boehler}, {Mace}, {Liu}, {Mulders}, {Hendler},
  {Ragusa}, {Fischer}, {Banzatti}, {Rigliaco}, {van de Plas}, {Dipierro},
  {Gully-Santiago}, \& {Lopez-Valdivia}}]{Long2019}
{Long}, F., {Herczeg}, G.~J., {Harsono}, D., {et~al.} 2019, \apj, 882, 49,
  \dodoi{10.3847/1538-4357/ab2d2d}

\bibitem[{{Maiwald} {et~al.}(2000){Maiwald}, {Lewen}, {Ahrens}, {Beaky},
  {Gendriesch}, {Koroliev}, {Negirev}, {Paveljev}, {Vowinkel}, \&
  {Winnewisser}}]{Maiwald2000}
{Maiwald}, F., {Lewen}, F., {Ahrens}, V., {et~al.} 2000, Journal of Molecular
  Spectroscopy, 202, 166, \dodoi{10.1006/jmsp.2000.8118}

\bibitem[{{Mart{\'\i}n-Dom{\'e}nech} {et~al.}(2019){Mart{\'\i}n-Dom{\'e}nech},
  {Bergner}, {{\"O}berg}, \& {J{\o}rgensen}}]{Martin-Domenech2019}
{Mart{\'\i}n-Dom{\'e}nech}, R., {Bergner}, J.~B., {{\"O}berg}, K.~I., \&
  {J{\o}rgensen}, J.~K. 2019, \apj, 880, 130, \dodoi{10.3847/1538-4357/ab2a08}

\bibitem[{{Marty} {et~al.}(2011){Marty}, {Chaussidon}, {Wiens}, {Jurewicz}, \&
  {Burnett}}]{Marty2011}
{Marty}, B., {Chaussidon}, M., {Wiens}, R.~C., {Jurewicz}, A.~J.~G., \&
  {Burnett}, D.~S. 2011, Science, 332, 1533, \dodoi{10.1126/science.1204656}

\bibitem[{{McClure} {et~al.}(2016){McClure}, {Bergin}, {Cleeves}, {van
  Dishoeck}, {Blake}, {Evans}, {Green}, {Henning}, {{\"O}berg}, {Pontoppidan},
  \& {Salyk}}]{McClure2016}
{McClure}, M.~K., {Bergin}, E.~A., {Cleeves}, L.~I., {et~al.} 2016, \apj, 831,
  167, \dodoi{10.3847/0004-637X/831/2/167}

\bibitem[{{Meier} {et~al.}(1998){Meier}, {Owen}, {Jewitt}, {Matthews}, {Senay},
  {Biver}, {Bockelee-Morvan}, {Crovisier}, \& {Gautier}}]{Meier1998}
{Meier}, R., {Owen}, T.~C., {Jewitt}, D.~C., {et~al.} 1998, Science, 279, 1707,
  \dodoi{10.1126/science.279.5357.1707}

\bibitem[{{Meijerink} {et~al.}(2009){Meijerink}, {Pontoppidan}, {Blake},
  {Poelman}, \& {Dullemond}}]{Meijerink2009}
{Meijerink}, R., {Pontoppidan}, K.~M., {Blake}, G.~A., {Poelman}, D.~R., \&
  {Dullemond}, C.~P. 2009, \apj, 704, 1471,
  \dodoi{10.1088/0004-637X/704/2/1471}

\bibitem[{{Messenger} \& {Walker}(1997)}]{Messenger1997}
{Messenger}, S., \& {Walker}, R.~M. 1997, in American Institute of Physics
  Conference Series, Vol. 402, American Institute of Physics Conference Series,
  ed. T.~J. {Bernatowicz} \& E.~{Zinner}, 545--564

\bibitem[{{Milam} {et~al.}(2005){Milam}, {Savage}, {Brewster}, {Ziurys}, \&
  {Wyckoff}}]{Milam2005}
{Milam}, S.~N., {Savage}, C., {Brewster}, M.~A., {Ziurys}, L.~M., \& {Wyckoff},
  S. 2005, \apj, 634, 1126, \dodoi{10.1086/497123}

\bibitem[{{Millar} {et~al.}(1989){Millar}, {Bennett}, \& {Herbst}}]{Millar1989}
{Millar}, T.~J., {Bennett}, A., \& {Herbst}, E. 1989, \apj, 340, 906,
  \dodoi{10.1086/167444}

\bibitem[{{Miotello} {et~al.}(2019){Miotello}, {Facchini}, {van Dishoeck},
  {Cazzoletti}, {Testi}, {Williams}, {Ansdell}, {van Terwisga}, \& {van der
  Marel}}]{Miotello2019}
{Miotello}, A., {Facchini}, S., {van Dishoeck}, E.~F., {et~al.} 2019, \aap,
  631, A69, \dodoi{10.1051/0004-6361/201935441}

\bibitem[{{M{\"u}ller} {et~al.}(2000){M{\"u}ller}, {Klaus}, \&
  {Winnewisser}}]{Muller2000}
{M{\"u}ller}, H.~S.~P., {Klaus}, T., \& {Winnewisser}, G. 2000, \aap, 357, L65

\bibitem[{{M{\"u}ller} {et~al.}(2005){M{\"u}ller}, {Schl{\"o}der}, {Stutzki},
  \& {Winnewisser}}]{Muller2005}
{M{\"u}ller}, H. S.~P., {Schl{\"o}der}, F., {Stutzki}, J., \& {Winnewisser}, G.
  2005, Journal of Molecular Structure, 742, 215,
  \dodoi{10.1016/j.molstruc.2005.01.027}

\bibitem[{{M{\"u}ller} {et~al.}(2001){M{\"u}ller}, {Thorwirth}, {Roth}, \&
  {Winnewisser}}]{Muller2001}
{M{\"u}ller}, H.~S.~P., {Thorwirth}, S., {Roth}, D.~A., \& {Winnewisser}, G.
  2001, \aap, 370, L49, \dodoi{10.1051/0004-6361:20010367}

\bibitem[{{Murillo} {et~al.}(2013){Murillo}, {Lai}, {Bruderer}, {Harsono}, \&
  {van Dishoeck}}]{Murillo2013}
{Murillo}, N.~M., {Lai}, S.-P., {Bruderer}, S., {Harsono}, D., \& {van
  Dishoeck}, E.~F. 2013, \aap, 560, A103, \dodoi{10.1051/0004-6361/201322537}

\bibitem[{{{\"O}berg} {et~al.}(2011){{\"O}berg}, {Murray-Clay}, \&
  {Bergin}}]{Oberg2011}
{{\"O}berg}, K.~I., {Murray-Clay}, R., \& {Bergin}, E.~A. 2011, \apjl, 743,
  L16, \dodoi{10.1088/2041-8205/743/1/L16}

\bibitem[{{Ohashi} {et~al.}(2014){Ohashi}, {Saigo}, {Aso}, {Aikawa},
  {Koyamatsu}, {Machida}, {Saito}, {Takahashi}, {Takakuwa}, {Tomida},
  {Tomisaka}, \& {Yen}}]{Ohashi2014}
{Ohashi}, N., {Saigo}, K., {Aso}, Y., {et~al.} 2014, \apj, 796, 131,
  \dodoi{10.1088/0004-637X/796/2/131}

\bibitem[{{Ortiz-Le{\'o}n} {et~al.}(2018){Ortiz-Le{\'o}n}, {Loinard}, {Dzib},
  {Kounkel}, {Galli}, {Tobin}, {Evans}, {Hartmann}, {Rodr{\'\i}guez},
  {Brice{\~n}o}, {Torres}, \& {Mioduszewski}}]{Ortiz2018}
{Ortiz-Le{\'o}n}, G.~N., {Loinard}, L., {Dzib}, S.~A., {et~al.} 2018, \apjl,
  869, L33, \dodoi{10.3847/2041-8213/aaf6ad}

\bibitem[{{Ossenkopf} \& {Henning}(1994)}]{Ossenkopf1994}
{Ossenkopf}, V., \& {Henning}, T. 1994, \aap, 291, 943

\bibitem[{{Oya} {et~al.}(2014){Oya}, {Sakai}, {Sakai}, {Watanabe}, {Hirota},
  {Lindberg}, {Bisschop}, {J{\o}rgensen}, {van Dishoeck}, \&
  {Yamamoto}}]{Oya2014}
{Oya}, Y., {Sakai}, N., {Sakai}, T., {et~al.} 2014, \apj, 795, 152,
  \dodoi{10.1088/0004-637X/795/2/152}

\bibitem[{{Padovani} {et~al.}(2009){Padovani}, {Walmsley}, {Tafalla}, {Galli},
  \& {M{\"u}ller}}]{Padovani2009}
{Padovani}, M., {Walmsley}, C.~M., {Tafalla}, M., {Galli}, D., \& {M{\"u}ller},
  H.~S.~P. 2009, \aap, 505, 1199, \dodoi{10.1051/0004-6361/200912547}

\bibitem[{{Pagani} {et~al.}(1992){Pagani}, {Salez}, \& {Wannier}}]{Pagani1992}
{Pagani}, L., {Salez}, M., \& {Wannier}, P.~G. 1992, \aap, 258, 479

\bibitem[{{Pegues} {et~al.}(2020){Pegues}, {{\"O}berg}, {Bergner}, {Loomis},
  {Qi}, {Gal}, {Cleeves}, {Guzm{\'a}n}, {Huang}, {J{\o}rgensen}, {Andrews},
  {Blake}, {Carpenter}, {Schwarz}, {Williams}, \& {Wilner}}]{Pegues2020}
{Pegues}, J., {{\"O}berg}, K.~I., {Bergner}, J.~B., {et~al.} 2020, \apj, 890,
  142, \dodoi{10.3847/1538-4357/ab64d9}

\bibitem[{{Qi} {et~al.}(2008){Qi}, {Wilner}, {Aikawa}, {Blake}, \&
  {Hogerheijde}}]{Qi2008}
{Qi}, C., {Wilner}, D.~J., {Aikawa}, Y., {Blake}, G.~A., \& {Hogerheijde},
  M.~R. 2008, \apj, 681, 1396, \dodoi{10.1086/588516}

\bibitem[{{Reboussin} {et~al.}(2015){Reboussin}, {Wakelam}, {Guilloteau},
  {Hersant}, \& {Dutrey}}]{Reboussin2015}
{Reboussin}, L., {Wakelam}, V., {Guilloteau}, S., {Hersant}, F., \& {Dutrey},
  A. 2015, \aap, 579, A82, \dodoi{10.1051/0004-6361/201525885}

\bibitem[{{Remijan} {et~al.}(2020){Remijan}, {Biggs}, {Cortes}, {Dent}, {Di
  Francesco}, {Fomalont}, {Hales}, {Kameno}, {Mason}, {Philips}, {Saini},
  {Stoehr}, {Vila Vilaro}, \& {Villard}}]{Remijan2020}
{Remijan}, A., {Biggs}, A., {Cortes}, P., {et~al.} 2020, ALMA Technical
  Handbook

\bibitem[{{Roberts} {et~al.}(2002){Roberts}, {Fuller}, {Millar}, {Hatchell}, \&
  {Buckle}}]{Roberts2002}
{Roberts}, H., {Fuller}, G.~A., {Millar}, T.~J., {Hatchell}, J., \& {Buckle},
  J.~V. 2002, \aap, 381, 1026, \dodoi{10.1051/0004-6361:20011596}

\bibitem[{{Roberts} \& {Millar}(2000)}]{Roberts2000}
{Roberts}, H., \& {Millar}, T.~J. 2000, \aap, 361, 388

\bibitem[{{Rodgers} \& {Charnley}(2008)}]{Rodgers2008}
{Rodgers}, S.~D., \& {Charnley}, S.~B. 2008, \apj, 689, 1448,
  \dodoi{10.1086/592195}

\bibitem[{{Rosenfeld} {et~al.}(2012){Rosenfeld}, {Andrews}, {Wilner}, \&
  {Stempels}}]{Rosenfeld2012}
{Rosenfeld}, K.~A., {Andrews}, S.~M., {Wilner}, D.~J., \& {Stempels}, H.~C.
  2012, \apj, 759, 119, \dodoi{10.1088/0004-637X/759/2/119}

\bibitem[{{Roueff} {et~al.}(2013){Roueff}, {Gerin}, {Lis}, {Wootten},
  {Marcelino}, {Cernicharo}, \& {Tercero}}]{Roueff2013}
{Roueff}, E., {Gerin}, M., {Lis}, D.~C., {et~al.} 2013, Journal of Physical
  Chemistry A, 117, 9959, \dodoi{10.1021/jp400119a}

\bibitem[{{Sastry} {et~al.}(1981){Sastry}, {Helminger}, {Charo}, {Herbst}, \&
  {De Lucia}}]{Sastry1981}
{Sastry}, K.~V.~L.~N., {Helminger}, P., {Charo}, A., {Herbst}, E., \& {De
  Lucia}, F.~C. 1981, \apjl, 251, L119, \dodoi{10.1086/183706}

\bibitem[{{Schwarz} {et~al.}(2016){Schwarz}, {Bergin}, {Cleeves}, {Blake},
  {Zhang}, {{\"O}berg}, {van Dishoeck}, \& {Qi}}]{Schwarz2016}
{Schwarz}, K.~R., {Bergin}, E.~A., {Cleeves}, L.~I., {et~al.} 2016, \apj, 823,
  91, \dodoi{10.3847/0004-637X/823/2/91}

\bibitem[{{Schwarz} {et~al.}(2018){Schwarz}, {Bergin}, {Cleeves}, {Zhang},
  {{\"O}berg}, {Blake}, \& {Anderson}}]{Schwarz2018}
---. 2018, \apj, 856, 85, \dodoi{10.3847/1538-4357/aaae08}

\bibitem[{{Schwarz} {et~al.}(2019){Schwarz}, {Bergin}, {Cleeves}, {Zhang},
  {{\"O}berg}, {Blake}, \& {Anderson}}]{Schwarz2019}
---. 2019, \apj, 877, 131, \dodoi{10.3847/1538-4357/ab1c5e}

\bibitem[{{Teague} {et~al.}(2015){Teague}, {Semenov}, {Guilloteau}, {Henning},
  {Dutrey}, {Wakelam}, {Chapillon}, \& {Pietu}}]{Teague2015}
{Teague}, R., {Semenov}, D., {Guilloteau}, S., {et~al.} 2015, \aap, 574, A137,
  \dodoi{10.1051/0004-6361/201425268}

\bibitem[{{Terzieva} \& {Herbst}(2000)}]{Terzieva2000}
{Terzieva}, R., \& {Herbst}, E. 2000, \mnras, 317, 563,
  \dodoi{10.1046/j.1365-8711.2000.03618.x}

\bibitem[{{Turner}(2001)}]{Turner2001}
{Turner}, B.~E. 2001, \apjs, 136, 579, \dodoi{10.1086/322536}

\bibitem[{{van der Walt} {et~al.}(2011){van der Walt}, {Colbert}, \&
  {Varoquaux}}]{VanDerWalt2011}
{van der Walt}, S., {Colbert}, S.~C., \& {Varoquaux}, G. 2011, Computing in
  Science and Engineering, 13, 22, \dodoi{10.1109/MCSE.2011.37}

\bibitem[{{van 't Hoff} {et~al.}(2018){van 't Hoff}, {Tobin}, {Harsono}, \&
  {van Dishoeck}}]{vantHoff2018}
{van 't Hoff}, M. L.~R., {Tobin}, J.~J., {Harsono}, D., \& {van Dishoeck},
  E.~F. 2018, \aap, 615, A83, \dodoi{10.1051/0004-6361/201732313}

\bibitem[{{Virtanen} {et~al.}(2020){Virtanen}, {Gommers}, {Oliphant},
  {Haberland}, {Reddy}, {Cournapeau}, {Burovski}, {Peterson}, {Weckesser},
  {Bright}, {van der Walt}, {Brett}, {Wilson}, {Jarrod Millman}, {Mayorov},
  {Nelson}, {Jones}, {Kern}, {Larson}, {Carey}, {Polat}, {Feng}, {Moore}, {Vand
  erPlas}, {Laxalde}, {Perktold}, {Cimrman}, {Henriksen}, {Quintero}, {Harris},
  {Archibald}, {Ribeiro}, {Pedregosa}, {van Mulbregt}, \&
  {Contributors}}]{SciPy2020}
{Virtanen}, P., {Gommers}, R., {Oliphant}, T.~E., {et~al.} 2020, Nature
  Methods, 17, 261, \dodoi{https://doi.org/10.1038/s41592-019-0686-2}

\bibitem[{{Visser} {et~al.}(2007){Visser}, {Geers}, {Dullemond}, {Augereau},
  {Pontoppidan}, \& {van Dishoeck}}]{Visser2007}
{Visser}, R., {Geers}, V.~C., {Dullemond}, C.~P., {et~al.} 2007, \aap, 466,
  229, \dodoi{10.1051/0004-6361:20066829}

\bibitem[{{Wampfler} {et~al.}(2014){Wampfler}, {J{\o}rgensen}, {Bizzarro}, \&
  {Bisschop}}]{Wampfler2014}
{Wampfler}, S.~F., {J{\o}rgensen}, J.~K., {Bizzarro}, M., \& {Bisschop}, S.~E.
  2014, \aap, 572, A24, \dodoi{10.1051/0004-6361/201423773}

\bibitem[{{Willacy}(2007)}]{Willacy2007}
{Willacy}, K. 2007, \apj, 660, 441, \dodoi{10.1086/512796}

\bibitem[{{Williams} \& {Best}(2014)}]{Williams2014}
{Williams}, J.~P., \& {Best}, W. M.~J. 2014, \apj, 788, 59,
  \dodoi{10.1088/0004-637X/788/1/59}

\bibitem[{{Wilson} \& {Rood}(1994)}]{Wilson1994}
{Wilson}, T.~L., \& {Rood}, R. 1994, \araa, 32, 191,
  \dodoi{10.1146/annurev.aa.32.090194.001203}

\bibitem[{{Winnewisser} {et~al.}(1997){Winnewisser}, {Belov}, {Klaus}, \&
  {Schieder}}]{Winnewisser1997}
{Winnewisser}, G., {Belov}, S.~P., {Klaus}, T., \& {Schieder}, R. 1997, Journal
  of Molecular Spectroscopy, 184, 468, \dodoi{10.1006/jmsp.1997.7341}

\bibitem[{{Winnewisser} {et~al.}(1985){Winnewisser}, {Winnewisser}, \&
  {Winnewisser}}]{Winnewisser1985}
{Winnewisser}, M., {Winnewisser}, B.~P., \& {Winnewisser}, G. 1985, in NATO
  Advanced Science Institutes (ASI) Series C, Vol. 157, NATO Advanced Science
  Institutes (ASI) Series C, ed. G.~H.~F. {Diercksen}, W.~F. {Huebner}, \&
  P.~W. {Langhoff}, 375--402

\bibitem[{{Zhang} {et~al.}(2017){Zhang}, {Bergin}, {Blake}, {Cleeves}, \&
  {Schwarz}}]{Zhang2017}
{Zhang}, K., {Bergin}, E.~A., {Blake}, G.~A., {Cleeves}, L.~I., \& {Schwarz},
  K.~R. 2017, Nature Astronomy, 1, 0130, \dodoi{10.1038/s41550-017-0130}

\bibitem[{{Zhang} {et~al.}(2019){Zhang}, {Bergin}, {Schwarz}, {Krijt}, \&
  {Ciesla}}]{Zhang2019}
{Zhang}, K., {Bergin}, E.~A., {Schwarz}, K., {Krijt}, S., \& {Ciesla}, F. 2019,
  \apj, 883, 98, \dodoi{10.3847/1538-4357/ab38b9}

\bibitem[{{Zhang} {et~al.}(2020){Zhang}, {Schwarz}, \& {Bergin}}]{Zhang2020}
{Zhang}, K., {Schwarz}, K.~R., \& {Bergin}, E.~A. 2020, \apjl, 891, L17,
  \dodoi{10.3847/2041-8213/ab7823}

\bibitem[{{Zhang} {et~al.}(2018){Zhang}, {Zhu}, {Huang}, {Guzm{\'a}n},
  {Andrews}, {Birnstiel}, {Dullemond}, {Carpenter}, {Isella}, {P{\'e}rez},
  {Benisty}, {Wilner}, {Baruteau}, {Bai}, \& {Ricci}}]{Zhang2018}
{Zhang}, S., {Zhu}, Z., {Huang}, J., {et~al.} 2018, \apjl, 869, L47,
  \dodoi{10.3847/2041-8213/aaf744}

\end{thebibliography}

\end{document}